\documentclass{aa}
\usepackage{graphics}
\usepackage{epsfig}
\usepackage{graphicx}
\usepackage{txfonts}
\usepackage{color}
\usepackage{subfigure}

\begin{document}

\color{black}

\title{An ISOCAM survey through gravitationally lensing galaxy clusters.
\thanks{Based on observations with ISO, an ESA project with instruments funded by ESA Member States
(especially the PI countries: France, Germany, the Netherlands and the United Kingdom) and with the participation
of ISAS and NASA. This work has also benefitted from ESO program I.D. 65.O-0489(A).}}

\subtitle{IV. Luminous infrared galaxies in Cl 0024+1654 and the dynamical status of clusters}

\titlerunning{An ISOCAM survey through gravitationally lensing galaxy clusters. IV.}

\author{D.~Coia \inst{1}       \and
        B.~McBreen\inst{1}     \and
        L.~Metcalfe\inst{2, 3} \and
        A.~Biviano\inst{4}     \and
        B.~Altieri \inst{2}    \and
        S.~Ott\inst{5}         \and
        B. Fort\inst{6}        \and
        J.-P.~Kneib\inst{7,8}  \and
        Y.~Mellier\inst{6,9}   \and
        M.-A. Miville-Desch\^{e}nes\inst{10} \and
        B.~O'Halloran\inst{1,11} \and
        C.~Sanchez-Fernandez\inst{2}.
        }

\authorrunning{Coia et~al.}

\offprints{D. Coia, \email{dcoia@bermuda.ucd.ie}}

\institute{     Department of Experimental Physics, University College, Belfield, Dublin 4, Ireland.
           \and XMM-Newton Science Operations Centre, European Space Agency, Villafranca del Castillo, P.O. Box 50727,
                28080 Madrid, Spain.
           \and ISO Data Centre, European Space Agency, Villafranca del Castillo, P.O. Box 50727, 28080 Madrid, Spain.
           \and INAF/Osservatorio Astronomico di Trieste, via G.B. Tiepolo 11, 34131, Trieste, Italy.
           \and Science Operations and Data Systems Division of ESA, ESTEC, Keplerlaan 1, 2200 AG Noordwijk,
                The Netherlands.
           \and Institut d'Astrophysique de Paris, 98 bis boulevard Arago, 75014 Paris, France.
           \and Observatoire Midi-Pyr\'en\'ees, 14 avenue Edouard Belin, 31400 Toulouse, France.
           \and California Institute of Technology, Pasadena, CA 91125, USA.
           \and Observatoire de Paris, 61 avenue de l'Observatoire, 75014 Paris, France.
           \and Canadian Institute for Theoretical Astrophysics, 60 St-George Street, Toronto, Ontario, M5S 3H8, Canada.
           \and Dunsink Observatory, Castleknock, Dublin 15, Ireland.
           }

\date{Received 30 July 2003 / Accepted 26 October 2004}

\abstract{ Observations of the core of the massive cluster  \object{Cl 0024+1654}, at a redshift z $\sim$ 0.39,
were obtained with the Infrared Space Observatory using ISOCAM at 6.7\,$\mu$m (hereafter 7\,$\mu$m) and
14.3\,$\mu$m (hereafter 15\,$\mu$m). Thirty five sources were detected at 15\,$\mu$m and thirteen of them are
spectroscopically identified with cluster galaxies. The remaining sources consist of four stars, one quasar, one
foreground galaxy, three background galaxies and thirteen sources with unknown redshift. The sources with
unknown redshift are all likely to be background sources that are gravitationally lensed by the cluster.

\hspace{4mm} The spectral energy distributions (SEDs) of twelve cluster galaxies were fit from a selection of 20
models using the program GRASIL.  The ISOCAM sources have best-fit SEDs typical of spiral or starburst models
observed 1 Gyr after the main starburst event. The star formation rates were obtained for cluster members. The
median infrared luminosity of the twelve cluster galaxies is $\sim1.0\times10^{11}$ $\mathrm{L}_\odot$, with 10 having
infrared luminosity above $9\times10^{10}$ $\mathrm{L}_\odot$, and so lying near or above the $1\times10^{11}$ $\mathrm{L}_\odot$
threshold for identification as a luminous infrared galaxy (LIRG). The [\ion{O}{ii}] star formation rates obtained for 3
cluster galaxies are one to two orders of magnitude lower than the infrared values, implying that most of the
star formation is missed in the optical because it is enshrouded by dust in the starburst galaxy.

\hspace{4mm}The cluster galaxies in general are spatially more concentrated than those
detected at 15\,$\mu$m. However the velocity distributions of the two categories are comparable. The
colour$-$magnitude diagramme is given for the galaxies within the ISOCAM map. Only 20\% of the galaxies that are
significantly bluer than the cluster main sequence were detected at 15\,$\mu$m, to the limiting sensitivity
recorded.  The counterparts of about half of the 15\,$\mu$m cluster sources are blue, luminous, star-forming
systems and the type of galaxy that is usually associated with the Butcher-Oemler effect. HST images of these
galaxies reveal a disturbed morphology with a tendency for an absence of nearby companions. Surprisingly the
counterparts of the remaining 15\,$\mu$m cluster galaxies lie on the main sequence of the colour-magnitude
diagramme. However in HST images they all have nearby companions and appear to be involved in interactions and
mergers. Dust obscuration may be a major cause of the 15\,$\mu$m sources appearing on the cluster main sequence.
The majority of the ISOCAM sources in the Butcher-Oemler region of the colour-magnitude diagram are best fit by
spiral-type SEDs whereas post-starburst models are preferred on the main sequence, with the
starburst event probably triggered by interaction with one or more galaxies.

\hspace{4mm}Finally, the mid-infrared results on Cl 0024+1654 are compared with four other clusters observed
with ISOCAM. Scaling the LIRG count in Cl 0024+1654 to the clusters  \object{Abell 370},  \object{Abell 1689},
 \object{Abell 2218} and \object{Abell 2390} with reference to their virial radii, masses, distances, and the sky area scanned in each case, we
compared the number of LIRGs observed in each cluster. The number in Abell 370 is smaller than expected by about
an order of magnitude, even though the two clusters are very similar in mass, redshift and optical richness. The
number of LIRGs detected in each of Abell 1689, Abell 2218 and Abell 2390 is 0, whereas 3 were expected from the
comparison with Cl 0024+1654. A comparison of the mid-infrared sources in Abell 1689 and Abell 2218 shows that
the sources in Abell 1689 are more luminous and follow the same trend identified in the comparison between Cl
0024+1654 and Abell 370. These trends seem to be related to the dynamical status and history of the clusters.
}

\maketitle

\section{Introduction\label{sec:sec1}}    %  SECTION 1

Clusters of galaxies contain thousands of members within a region a few Mpc in diameter, and are the largest
known gravitationally bound systems of galaxies, having masses up to $10^{15}$ $\mathrm{M}_\odot$ for the richest systems.
In hierarchical models clusters of galaxies grow by accreting less massive groups falling along filaments at a
rate governed by the initial density fluctuation spectrum, the cosmological parameters and the nature and amount
of dark matter. In the cluster environment, newly added galaxies are transformed from blue, active star forming
systems, to red, passive ellipticals, undergoing a morphological evolution stronger than that of field galaxies
at a similar redshift (Gavazzi \& Jaffe \cite{1987A&A...186L...1G}; Byrd \& Valtonen \cite{1990ApJ...350...89B};
Abraham et al. \cite{1996ApJ...471..694A}). The cluster galaxy population is also characterized by a lower star
formation rate (SFR) than field galaxies of similar physical size and redshift (Couch et al.
\cite{2001ApJ...549..820C}; Lewis et al. \cite{2002MNRAS.334..673L}).

Butcher and Oemler (\cite{1978ApJ...219...18B}) showed that clusters of galaxies generally have a fraction
f$_\mathrm{B}$ of blue galaxies\footnote{Blue galaxies are defined as brighter than M$_\mathrm{V}=-19.26$ (with
H$_0=70\,\mathrm{km}\,\mathrm{s}^{-1}\,\mathrm{Mpc}^{-1}$) with rest-frame B-V colours at least 0.2 magnitudes
bluer than those of the E/S0 galaxy sequence at the same absolute magnitude (Butcher \& Oemler
\cite{1984ApJ...285..426B}; Oemler et al. \cite{1997ApJ...474..561O}).} that increases with cluster redshift,
ranging from a value near 0 at $z = 0$, to 20\% at $z = 0.4$ and to 80\% at $\mathrm{z} = 0.9$, suggesting a
strong evolution in clusters (Rakos \& Schombert \cite{1995ApJ...439...47R}). The galaxies responsible for the
Butcher-Oemler effect (hereafter BO effect) are generally luminous, spirals, and emission-line systems, with
disturbed morphologies. The high resolution imaging achieved by the Hubble Space Telescope (HST) has greatly
improved morphological studies of the cluster galaxy population over a wide range in redshift. The mixture of
Hubble types in distant clusters is significantly different from that seen in nearby systems. The population of
star-forming and post-starburst galaxies are disk dominated systems, some of which are involved in interactions
and mergers. (Abraham et al \cite{1996ApJS..107....1A}; Stanford et al. \cite{1998ApJ...492..461S}; Couch et al.
\cite{1998ApJ...497..188C}; Van Dokkum et al. \cite{1998ApJ...500..714V}; Morris et al.
\cite{1998ApJ...507...84M}; Poggianti et al. \cite{1999ApJ...518..576P}; Best \cite{2000MNRAS.317..720B}).

Many mechanisms have been proposed to explain the complicated processes that occur in clusters, including ram
pressure stripping of gas (Gunn \& Gott \cite{1972ApJ...176....1G}), galaxy harassment (Moore et al.
\cite{1996Natur.379..613M}; Moss \& Whittle \cite{1997RMxAC...6..145M}), galaxy infall (Ellingson et al.
\cite{2001ApJ...547..609E}), cluster tidal forces (Byrd \& Valtonen \cite{1990ApJ...350...89B}; Fujita
\cite{1998ApJ...509..587F}) and interactions with other cluster galaxies (Icke \cite{1985A&A...144..115I}; Moss
\& Whittle \cite{1997RMxAC...6..145M}). The main processes responsible for the morphological and spectral
evolution of cluster galaxies have yet to be determined. Ram pressure and tidal effects can quench the star
formation activity gradually because they operate over a period longer than 1 Gyr (e.g. Ghigna et al.
\cite{1998MNRAS.300..146G}; Ramirez \& de Souza \cite{1998ApJ...496..693R}). Galaxy-galaxy interactions, galaxy
harassment and cluster mergers can enhance it and produce changes in galaxy properties over timescales of
$\sim100$ Myr (e.g. Lavery \& Henry \cite{1986ApJ...304L...5L}; Moore et~al. \cite{1996Natur.379..613M}). Recent
changes in the properties of the galaxies may be detectable in the mid-infrared if associated with a burst of
star formation.

In the context of our ongoing exploitation of mid-infrared cluster data obtained with ESA's Infrared Space
Observatory (ISO, Kessler et al. \cite{1996A&A...315L..27K}), we have analysed ISO observations of Abell 370,
Abell 2218 \& Abell 2390 (Metcalfe et al. \cite{2003A&A...407..791M}; Altieri et al. \cite{1999A&A...343L..65A};
Biviano et al. \cite{biviano04}), Abell 2219 (Coia et al. \cite{coia04}), and Cl 0024+1654 (this paper).  In
common with other surveys we have found that mid-infrared observations of field galaxies from deep surveys
reveal a population of starburst galaxies that evolve significantly with redshift (Aussel  et al.
\cite{1999AAS...195.0917A}; Oliver et al. \cite{2000MNRAS.316..749O}; Serjeant et al.
\cite{2000MNRAS.316..768S}; Lari et al. \cite{2001MNRAS.325.1173L}; Gruppioni et al. \cite{2002MNRAS.335..831G};
Elbaz \& Cesarsky \cite{2003Sci...300..270E}; Metcalfe et al. \cite{2003A&A...407..791M}; Sato et al.
\cite{2003A&A...405..833S}). This class of sources are Luminous and Ultraluminous infrared galaxies (LIRGs and
ULIRGs, Sanders \& Mirabel \cite{1996ARA&A..34..749S}; Genzel \& Cesarsky \cite{2000ARA&A..38..761G}), have SFRs
of $\sim100$ $\mathrm{M}_\odot\,\mathrm{yr}^{-1}$ (Oliver et al. \cite{2000MNRAS.316..749O}; Mann et al. \cite{2002MNRAS.332..549M}) and seem
to be almost always the result of galaxy-galaxy interactions at least in the local Universe (Veilleux et al.
\cite{2002ApJS..143..315V}). The ISOCAM sources account for most of the contribution of the mid-infrared to the
Cosmic Infrared Background (CIRB, Altieri et al. \cite{1999A&A...343L..65A}; Franceschini et al.
\cite{2001A&A...378....1F}; Metcalfe et al. \cite{2001IAUS..204..217M,2003A&A...407..791M}; Elbaz et al.
\cite{2002A&A...384..848E}). Studies of the global SFR show a decline by a factor of $3-10$ since the peak of
star formation at $z= 1-2$ (Madau et al. \cite{1996MNRAS.283.1388M}; Steidel et al. \cite{1999ApJ...519....1S};
Elbaz \& Cesarsky \cite{2003Sci...300..270E}). The downturn in the global SFR and population of LIRGs and ULIRGs
may be caused by galaxies running out of gas available for star formation and the buildup of large scale
structure in the Universe that changed the environment of galaxies.

A study of the impact of the environment on galaxies in clusters could help in understanding the global SFR.
\begin{figure*}         % FIGURE 1
\centering
%\includegraphics[width=17cm]{green_may.eps}
%\includegraphics{overlay.ps}
%\resizebox{\hsize}{!}{\includegraphics{15mic_overlay_square_full_4.ps}}
\resizebox{2\columnwidth}{!}{\includegraphics{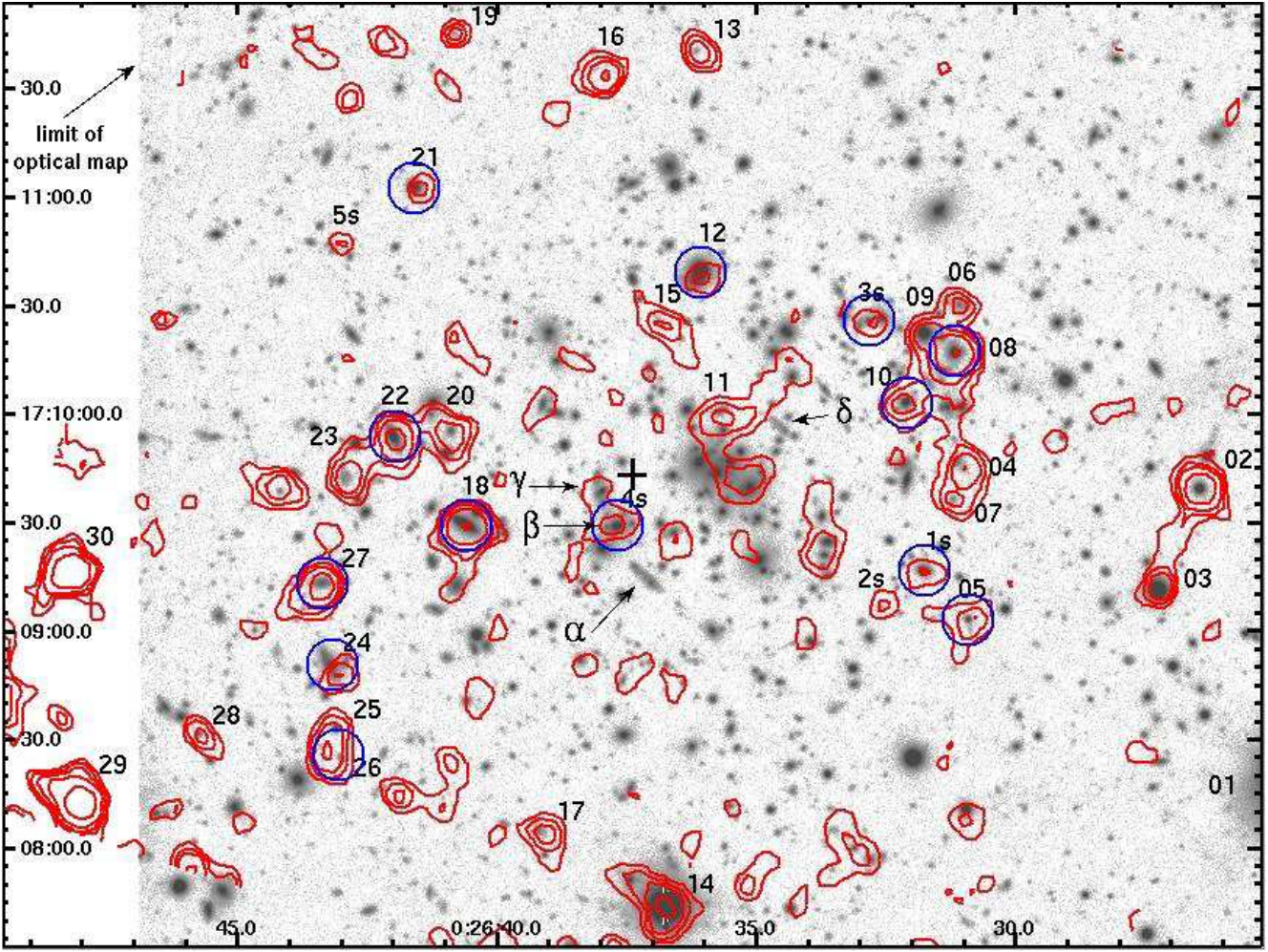}}\caption{The 15\,$\mu$m contour map (red) of Cl
0024+1654 overlaid on a Very Large Telescope image taken in the V band with the FORS2 instrument (ESO program
identification: 65.O-0489(A)). Numbers 1 to 30 refer to the 15\,$\mu$m sources in the primary list
(Table~\ref{lw3}) and sources 1s to 5s label sources from the supplementary list (Table~\ref{lw3_add}). Each set
of sources is labelled in order of increasing Right Ascension. Sources ISO\_Cl0024\_29 and ISO\_Cl0024\_30 are
outside the boundary of the optical map, and have a star and a faint galaxy, respectively, as optical
counterparts. Blue circles denote 15\,$\mu$m sources that are spectroscopically confirmed cluster galaxies.
Greek letters ($\alpha \div \delta$) identify four gravitationally lensed images of the background galaxy
associated with the spectacular giant arcs in the cluster field. North is up and East is to the left. The centre
of the ISO map, indicated by a cross, is at R.A. 00 26 37.5 and DEC. 17 09 43.4 (J2000). } \label{lw3_fig}
\end{figure*}
Mid-infrared observations have been published for local clusters (Boselli et al.
\cite{1997A&A...324L..13B,1998A&A...335...53B}; Contursi et al. \cite{2001A&A...365...11C}) and distant clusters
of galaxies (Pierre et al. \cite{1996A&A...315L.297P}; L{\' e}monon et al. \cite{1998A&A...334L..21L}; Altieri
et al. \cite{1999A&A...343L..65A}; Fadda et al. \cite{2000A&A...361..827F}; Metcalfe et al.
\cite{2003A&A...407..791M}; Coia et al. \cite{coia04}). In Abell 2390, Abell 370 and Abell 2218 the greater part
of the 15\,$\mu$m sources for which spectroscopic redshifts are available are found to be background sources
(Metcalfe et al. \cite{2003A&A...407..791M}).  However Fadda et al. (\cite{2000A&A...361..827F}) and Duc et al.
(\cite{2002A&A...382...60D}) found a higher proportion of 15\,$\mu$m cluster sources in the cluster Abell 1689
at $z = 0.18$ and Duc et al. (\cite{astro-ph/0404183}) discovered many LIRGs in the cluster J1888.16CL at
$z=0.56$.

In this work we focus on the mid-infrared properties of the galaxy cluster Cl 0024+1654. The paper is organized
as follows: Sect.~\ref{sec:cluster} contains a description of the cluster. Section~\ref{sec:obsda} describes the
infrared observations and outlines the data reduction, source extraction and calibration processes.
Section~\ref{sec:results} presents the results, the model spectral energy distributions (SEDs) and star
formation rates for cluster galaxies. Section~\ref{sec:distribution} describes the spatial, redshift and colour
properties of cluster galaxies and contains a description of Hubble Space Telescope images of some galaxies
detected by ISOCAM. Section~\ref{sec:compa} makes a comparison between Cl 0024+1654 and other clusters studied,
including Abell 1689, Abell 370, Abell 2390 and Abell 2218. The conclusions are in Sect.~\ref{sec:concl}. The
Appendix contains additional comments on some of the ISOCAM sources.

We adopt H$_0=70$\,km\,s$^{-1}$\,Mpc$^{-1}$, $\Omega_\Lambda=0.7$ and $\Omega_\mathrm{m}=0.3$. With this cosmology, the
luminosity
distance to the cluster is $\mathrm{D}_\mathrm{L}$ = 2140 Mpc and 1\arcsec\ corresponds to 5.3 kpc at the
cluster redshift. The age of the Universe at the cluster redshift of 0.39 is 9.3 Gyr.

\section{The cluster\label{sec:cluster}}   % SECTION 2

Cl 0024+1654 is a rich cluster of galaxies at redshift $z \sim 0.395$ (Humason \& Sandage \cite{humason}; Gunn
\& Oke \cite{1975ApJ...195..255G}; Smail et al. \cite{1993MNRAS.263..628S}). It has a spectacular system of
gravitationally lensed arcs (Fig.~\ref{lw3_fig}) that were first observed by Koo (\cite{1988lsmu.book..513K})
and subsequently studied by Mellier et al. (\cite{1991ApJ...380..334M}), Kassiola et al.
(\cite{1994ApJ...429L...9K}), Wallington et al. (\cite{1995ApJ...441...58W}) and Smail et al.
(\cite{1997ApJS..110..213S}). The main arc is split into segments (Colley et~al. \cite{1996ApJ...461L..83C};
Tyson et~al. \cite{1998ApJ...498L.107T}) and is the lensed image of a background blue galaxy at redshift $z =
1.675$ (Broadhurst et al. \cite{2000ApJ...534L..15B}). Unlike other clusters of galaxies, such as Abell 2218 or
Abell 2390 that have arc-like features, Cl 0024+1654 does not have a central dominant cD galaxy.

\begin{figure}%[!hb\textwidth]      %FIGURE 2
%\resizebox{\hsize}{!}{\includegraphics{histovc_new.eps}}
\resizebox{\hsize}{!}{\includegraphics{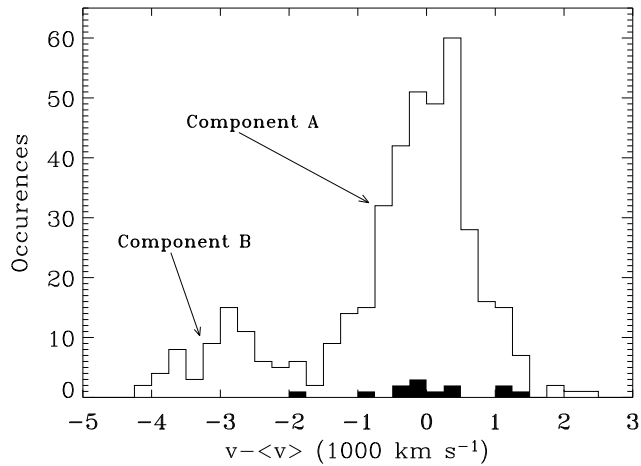}}\caption{The fractional distribution of velocities of
cluster galaxies in Cl 0024+1654. The counterparts of the cluster members detected at 15\,$\mu$m are displayed
in bold. All velocities are in the cluster rest-frame. } \label{fig:vel_disp}
\end{figure}
Cl 0024+1654 was one of the two clusters examined by Butcher \& Oemler (\cite{1978ApJ...219...18B}) in their
first published work on galaxy colours. The fraction of blue galaxies in Cl 0024+1654 is f$_\mathrm{B}$ = 0.16
which is much larger than the values of f$_\mathrm{B}$ = 0.04 and f$_\mathrm{B}$ = 0.03 for the Virgo ($z =
0.003$) and Coma ($z = 0.02$) clusters respectively (Butcher \& Oemler \cite{1984ApJ...285..426B}; Dressler et
al. \cite{1985ApJ...294...70D}; Schneider et al. \cite{1986AJ.....92..523S}).

The determination of the spectral types of galaxies has been made for the cluster and the field (Czoske et al.
\cite{2002A&A...386...31C}; Balogh et al. \cite{2002MNRAS.335...10B}). Using a wide field HST survey of Cl
0024+1654, Treu et al. (\cite{2003ApJ...591...53T}) found that the fraction of early-type galaxies (E+S0) is
highest ($\sim 73\%$) in the cluster core, declines rapidly to about 50\% at $\sim$ 1 Mpc, and reaches the
background value of $\sim\,43\%$ at the periphery of the cluster (at a radius of $\sim$ 5 Mpc).

The mass profile of the cluster has been inferred from gravitational lensing analyses, kinematical analyses of
the redshifts of the cluster galaxies, and X-ray observations. The low value of the X-ray luminosity implies a
mass 2 to 3 times smaller than that predicted by lensing models (Soucail et al. \cite{2000A&A...355..433S}; Ota
et al. \cite{2004ApJ...601..120O}). There are also differences among the lensing models. Broadhurst et al.
(\cite{2000ApJ...534L..15B}) favour a cuspy NFW (Navarro, Frenk \& White \cite{1997ApJ...490..493N}) profile for
the mass distribution of the cluster. Tyson et~al. (\cite{1998ApJ...498L.107T}) argue that this result requires
an average cluster velocity distribution much higher than the measured value of $\sigma_\mathrm{v}\sim
1150$\,km\,s$^{-1}$ (Dressler et al. \cite{1999ApJS..122...51D}) and favour a uniform-density core for the halo
mass profile. The weak lensing analysis made by Kneib et al. (\cite{2003ApJ...598..804K}), based on an extensive
HST survey of Cl 0024+1654 (Treu et al. \cite{2003ApJ...591...53T}), indicates that the region within 5 Mpc from
the centre of the cluster is well fit by a steep NFW-like profile, and the isothermal fit is strongly rejected.
Ota et~al. (\cite{2004ApJ...601..120O}) analysed the X-ray emission from the cluster and found that an
isothermal model with a mean temperature of 4.4 keV is a good fit to the temperature distribution. They also
estimated the mass of the cluster on the assumption that the intracluster medium is in hydrostatic equilibrium,
and confirmed that the mass inferred from X-ray observations is smaller than the mass predicted by lensing
models by a factor of about 3. The complex structure of the central region of the cluster might explain the
discrepancy (Zhang et al. \cite{astro-ph/0408545}).

During an extensive spectroscopic survey of Cl 0024+1654, Czoske et~al. (\cite{2001A&A...372..391C}) detected a
significant number of galaxies at a redshift slightly lower than that of the cluster (Fig.~\ref{fig:vel_disp}).
This concentration (component B) is interpreted as a small, less massive cluster superimposed on the main
cluster (component A) and lying at a mean redshift of $z = 0.381$. Numerical simulations indicate that the two
clusters in Cl 0024+1654 are involved in a high-speed collision along the line-of-sight to the cluster (Czoske
et al. \cite{2002A&A...386...31C}). The relative velocity of the two components, derived from the difference in
redshift of the two main galaxy concentrations, is about 3000 km\,s$^{-1}$. The collision scenario is not
unusual because 30-40\% of galaxy clusters have substructures that are detected in optical, X-ray and radio
observations (e.g. Dressler \& Shectman \cite{1988AJ.....95..985D}; Schuecker et al. \cite{2001A&A...378..408S};
Mercurio et al. \cite{2003A&A...397..431M}). These substructures indicate that the hosting clusters are not
fully dynamically relaxed and might be currently undergoing, or have recently undergone, mergers. The important
results obtained by Czoske et al. (\cite{2001A&A...372..391C,2002A&A...386...31C}) may give a new insight into
the BO effect in medium and high redshift clusters. The impact of the two clusters could have triggered star
formation in cluster galaxies, especially those on the leading edge of the smaller cluster, thus generating the
large number of blue galaxies observed in Cl 0024+1654. The differences between the results from the methods
used to determine the mass of Cl 0024+1654 are naturally explained by the cluster collision (Czoske et
al.~\cite{2002A&A...386...31C}).

\begin{table*}          %       TABLE 1
\caption[]{Observational parameters used. The observations were made with the ISOCAM LW2 (5 - 8.5\,$\mu$m) and
LW3 (12 - 18\,$\mu$m) filters at their respective reference wavelengths of 6.7\,$\mu$m and 14.3\,$\mu$m. On-chip
integration time was always 5.04 seconds and the 3\arcsec~per-pixel-field-of-view was used. M and N are the
number of steps along each dimension of the raster, while dm and dn are the increments for a raster step. The
sensitivity in $\mu$Jy is given at the 5$\sigma$ level. The table also includes the total area covered and the
number of readouts per raster step. The rasters were repeated $k$ times.  Tot. t is the total time dedicated to
each filter observation. \label{tab:isoobs} } \centering
\begin{tabular}{ccccccccccc}
       \hline\hline
\noalign{\smallskip}
Filter & $\lambda_\mathrm{ref}$&\multicolumn{2}{c}{n Steps}&dm       &dn       &Reads & area &Done k  &Sensitivity& Tot. t     \\
       &($\mu$m)& M & N                    &(\arcsec)&(\arcsec)&per step&$(\arcmin^2)$&times   & ($\mu$Jy)&(sec)      \\
       \hline
\noalign{\smallskip}
LW2    &6.7  &6  & 6  &45        &45      & 14     &  28.6   & 1     &400 &3138         \\
\noalign{\smallskip}
LW3    &14.3 &14 & 14 &21        &21      & 10     &  37.8   & 2     &140 &22615        \\
        \hline
      \end{tabular}
\end{table*}
\section{Observations, data reduction and source detection \label{sec:obsda}}   % SECTION 3
\subsection{Observations} %SECTION 3.1

The core of the cluster Cl 0024+1654 was observed at 7\,$\mu$m and 15\,$\mu$m using the LW2 and LW3 filters of
the camera ISOCAM (Cesarsky et al. \cite{1996A&A...315L..32C}) on board ISO. The ISOCAM long wavelength (LW)
detector consisted of a $32\times32$\ SiGa array. The observations were made on June 9 and June 15, 1997, in
raster mode, with an on-chip integration time of 5.04 seconds in the 3\arcsec~per pixel field of view, and cover
an area of approximately 38 square arcminutes. A discussion of the relevant observing strategy for ISOCAM can be
found in Metcalfe et al. (\cite{2003A&A...407..791M}) (with the exception that the Cl 0024+1654 rasters were not
``micro-scanned" with sub-pixel finesse. I.e. the raster step sizes were multiples of the array pixel size,
whereas Metcalfe et al. describe rasters with step sizes involving fractions of a pixel's dimensions). For each
pointing of the raster, 14 readouts were performed at 7\,$\mu$m and 10 at 15\,$\mu$m. Fifty readouts were taken
at the beginning of each raster to allow for detector stabilization. The parameters used for the observations
are given in Table~\ref{tab:isoobs}, which also lists the 5$\sigma$ sensitivity reached after data processing.
The diameter of the point spread function (PSF) central maximum at the first Airy minimum is
$0.84\times\lambda(\mu$m) arcseconds. The FWHM is about half that amount and Okumura (\cite{okumura}) obtained
values of 3.3\arcsec~at 7\,$\mu$m and 5\arcsec~at 15\,$\mu$m for the PSF FWHM in the 3\arcsec~per pixel
field-of-view. The data were reduced and, to take advantage of a slight deliberate offset between the two
15\,$\mu$m maps, were rebinned so that the final map has a pixel size of 1\arcsec, with potential for slightly
improved spatial resolution. The maximum depth is reached toward the centre of the rasters, where the dwell time
per position on the sky is greatest. As recorded in Table~\ref{tab:isoobs}, two rasters were made at 15\,$\mu$m
and a single raster at 7\,$\mu$m that had a larger step size and less observation time (by a factor of 7).

\subsection{Data reduction} %SECTION 3.2

The data were reduced using the {\em ISO{\bf C}AM {\bf I}nteractive {\bf A}nalysis} System or CIA (Delaney \&
Ott \cite{ISOCAM}; Ott et~al. \cite{1997adass...6...34O}) in conjunction with dedicated routines, following the
method\footnote{An interesting alternative approach, optimised for extended-source detection, but which gives
very good overall results and can be used to cross-check the validity of the faintest detections, is the method
of Miville-Desch\^{e}nes et al. (\cite{2000A&AS..146..519M}).  We employed products of this alternative
reduction process (SLICE) as a further cross-check of the reality of the faintest sources in our lists,
reasoning that both methods would tend to preserve real sources, while any residual glitches affecting the
results of our primary analysis would have some probability of being filtered out in the independent SLICE
analysis. To arrive at our final list we rejected some very faint candidate sources as being unreliable when
they were poorly reproduced by this independent analysis route. We consider this approach to be very
conservative. } described in Metcalfe et~al. (\cite{2003A&A...407..791M}). The two 15\,$\mu$m raster maps were
normalized to their respective redundancy maps\footnote{The ``redundancy" of a point in a raster map refers to
the number of raster steps for which that point on the sky has been seen by some detector pixel. } and merged
into a single raster, thus increasing the sensitivity of the map to faint sources.

The 15\,$\mu$m sources were found by visually inspecting the merged and the two individual maps. Detections
common to the three maps are given in the \textsl{primary source list} (Table~\ref{lw3}). The selection
criterion adopted is very conservative because it requires the source to be detected in both individual rasters,
resulting in the rejection of faint sources that can be detected only in the merged map. Therefore we include a
\textsl{supplementary list} containing sources found in the merged map and in at least one of the individual
rasters (Table~\ref{lw3_add}). The fluxes of the sources were computed by aperture photometry using a circular
aperture of 9\arcsec\ diameter, centred on the infrared source.

Since there was only one raster at 7\,$\mu$m, an independent comparison of source detections was not possible
and real sources were considered to be only those having a 15\,$\mu$m counterpart.

The optical data were obtained on the ESO/VLT during October 2001 with the VLT/FORS instrument in service
mode\footnote{Program ID: 65.0-489A FORS; PI: Fort}. The data have been processed at the TERAPIX data
center\footnote{{\tt http://terapix.iap.fr}}. Pre-calibrations, astrometric and photometric calibrations as well
as image stacking and catalog production were done using standard CCD image processing algorithms and tools
available at the TERAPIX center (see McCracken et al. \cite{2003A&A...410...17M} for details).
\begin{table*}                  % TABLE 2
\centering \caption[]{The primary list of sources detected at 15\,$\mu$m. The columns are (from left to right):
source identification number in order of increasing Right Ascension; source flux and precision before
calibration (in ADUgs); source flux and precision after calibration (in mJy); Right Ascension and Declination of
the infrared source (J2000); redshift and name of the optical counterpart, if known. The redshift of source
ISO\_Cl0024\_02 was taken from Schmidt et~al. (\cite{1986ApJ...306..411S}). All other redshifts are from Czoske
et al. (\cite{2001A&A...372..391C}) or provided by T. Treu (private communication). The names of the stars are
from HEASARC.\label{lw3}}
     \begin{tabular}{cccccccccc}
         \hline
         \hline
        \noalign{\smallskip}
Source ID  & Signal & Precision   &  Flux &  Precision  &   R.A.     &   DEC.      &   z      & Name of optical     \\
ISO\_Cl0024& (ADU)  & ($\pm$ ADU) & (mJy) & ($\pm$ mJy) & (J2000)    & (J2000)     & if known & counterpart if known\\
\hline
\noalign{\smallskip}
01&    0.634 &0.090 &   0.786&  0.120&    00 26 24.3 & +17 08 18.0&  Star  & TYC 1180-82-1  \\
02&    0.472 &0.060 &   0.577&  0.090&    00 26 26.3 & +17 09 39.0&  0.959 & PC 0023+1653   \\
03&    0.256 &0.060 &   0.298&  0.080&    00 26 27.0 & +17 09 11.7&  Star  & N3231320329    \\
04&    0.298 &0.060 &   0.352&  0.080&    00 26 30.8 & +17 09 44.8&  -     &                \\
05&    0.184 &0.060 &   0.205&  0.080&    00 26 30.7 & +17 09 03.5&  0.3935&                \\
06&    0.230 &0.060 &   0.264&  0.080&    00 26 30.9 & +17 10 29.4&  -     &                \\
07&    0.217 &0.060 &   0.247&  0.080&    00 26 31.0 & +17 09 37.4&  -     &                \\
08&    0.610 &0.090 &   0.755&  0.120&    00 26 31.0 & +17 10 15.6&  0.4005&                \\
09&    0.286 &0.060 &   0.336&  0.080&    00 26 31.5 & +17 10 21.4&  0.2132&                \\
10&    0.191 &0.060 &   0.214&  0.080&    00 26 31.8 & +17 10 03.5&  0.4000&                \\
11&    0.220 &0.060 &   0.251&  0.080&    00 26 35.5 & +17 09 59.1&  0.5558&                \\
12&    0.182 &0.060 &   0.202&  0.080&    00 26 35.9 & +17 10 36.4&  0.3860&                \\
13&    0.210 &0.060 &   0.238&  0.080&    00 26 35.9 & +17 11 39.9&  -     &                \\
14&    0.293 &0.060 &   0.346&  0.080&    00 26 36.6 & +17 07 44.4&  Star  &  EO903-0219757 \\
15&    0.170 &0.060 &   0.186&  0.080&    00 26 36.6 & +17 10 25.4&  -     &                \\
16&    0.260 &0.060 &   0.302&  0.080&    00 26 37.7 & +17 11 33.4&  -     &                \\
17&    0.253 &0.060 &   0.293&  0.080&    00 26 38.9 & +17 08 03.9&  -     &                \\
18&    0.634 &0.090 &   0.786&  0.120&    00 26 40.4 & +17 09 28.7&  0.394 &                \\
19&    0.187 &0.060 &   0.208&  0.080&    00 26 40.6 & +17 11 44.9&  -     &                \\
20&    0.262 &0.060 &   0.305&  0.080&    00 26 40.7 & +17 09 54.0&  0.7125&                \\
21&    0.191 &0.060 &   0.213&  0.080&    00 26 41.3 & +17 11 01.4&  0.3924&                \\
22&    0.334 &0.060 &   0.398&  0.080&    00 26 41.8 & +17 09 52.9&  0.3935&                \\
23&    0.282 &0.060 &   0.331&  0.080&    00 26 42.7 & +17 09 41.6&  -     &                \\
24&    0.189 &0.060 &   0.211&  0.080&    00 26 42.8 & +17 08 48.5&  0.3954&                \\
25&    0.200 &0.060 &   0.225&  0.080&    00 26 42.9 & +17 08 31.7&  0.9174&                \\
26&    0.324 &0.060 &   0.385&  0.080&    00 26 43.1 & +17 08 25.1&  0.3961&                \\
27&    0.373 &0.060 &   0.449&  0.080&    00 26 43.2 & +17 09 11.7&  0.3932&                \\
28&    0.120 &0.060 &   0.122&  0.080&    00 26 45.5 & +17 08 30.8&  -     &                \\
29&    0.946 &0.110 &   1.191&  0.140&    00 26 47.9 & +17 08 12.0&  Star  &  EO903-0219375 \\
30&    0.489 &0.070 &   0.598&  0.090&    00 26 48.0 & +17 09 15.7&  -     &                \\
\hline
      \end{tabular}
\end{table*}

\begin{table*}             % TABLE 3
\centering \caption[]{The supplementary list of 15\,$\mu$m sources. The meaning of the columns is the same as in
Table~\ref{lw3}. \label{lw3_add}}
   \begin{tabular}{ccccccccc}
         \hline
         \hline
Source ID        & Signal      & Precision   &  Flux       &  Precision  &    R.A.          &   DEC.            &  z           \\
ISO\_Cl0024 & (ADU)       & ($\pm$ ADU) & (mJy)       & ($\pm$ mJy) &  (J2000)         & (J2000)           & if
known     \\\hline \noalign{\smallskip}
1s &     0.150 &0.060 &   0.160&  0.080& 00 26 31.6 & +17 09 17.1 &  0.3998&          \\
2s &     0.190 &0.060 &   0.212&  0.080& 00 26 32.4 & +17 09 06.5 &  -       &        \\
3s &     0.124 &0.060 &   0.127&  0.080& 00 26 32.6 & +17 10 26.1 &  0.3965&          \\
4s &     0.173 &0.060 &   0.190&  0.080& 00 26 37.4 & +17 09 29.1 &  0.3900&          \\
5s &     0.182 &0.060 &   0.202&  0.080& 00 26 42.9 & +17 10 45.6 &  -       &        \\
\hline
      \end{tabular}
\end{table*}

\begin{table*}             % TABLE 4
   \centering
\caption{The list of 7\,$\mu$m sources with source flux exceeding 5$\sigma$ of the local noise, and having
15\,$\mu$m counterparts. From left to right: 15\,$\mu$m source identification number as listed in
Table~\ref{lw3}; signal and precision of flux in ADU; flux and precision in mJy; Right Ascension, Declination;
nature of the source; and finally the [7\,$\mu$m]/[15\,$\mu$m] colour ratio. \label{lw2}}
       \begin{tabular}{ccccccccccc}
\hline
         \hline
Source ID  &  Signal           &Precision             &  Flux         &  Precision           &   R.A.           &    DEC.          & Notes          & {\small [7\,$\mu$m]/[15\,$\mu$m]}\\
           & {\small (ADU)}    & {\small ($\pm$ ADU)} & {\small(mJy)} & {\small ($\pm$ mJy)} & {\small (J2000)} & {\small (J2000)} &                  &                                  \\
\hline \noalign{\smallskip}
03         & 1.461             & 0.219                & 1.685         & 0.253                & 00 26 27.0       &  +17 09 08.2     &   Star             &    5.7                           \\
14         & 1.276             & 0.191                & 1.468         & 0.220                & 00 26 36.5       &  +17 07 41.0     &   Star             &    4.2                          \\
29         & 3.806             & 0.571                & 4.427         & 0.664                & 00 26 47.9       &  +17 08 11.1     &   Star             &    3.7                          \\
\hline
       \end{tabular}
\end{table*}

\subsection{Monte Carlo simulations and calibration\label{sec:montecarlo} } %SECTION 3.3

Monte Carlo simulations were performed to calibrate the complex data reduction process by characterizing how it
affects model point sources with known properties inserted into the raw data, and to characterize the precision
with which source signals could be recovered from the data. The procedure adopted is fully described in Metcalfe
et~al. (\cite{2003A&A...407..791M}), the only exception being that all photometry performed for the present work
employed the XPHOT routine in CIA, whereas Metcalfe et~al. used SExtractor (Bertin \&
Arnouts~\cite{1996A&AS..117..393B}). The fluxes of the inserted fake sources were measured in exactly the same
way as the fluxes of the real sources. The simulations were performed independently for all rasters. The fluxes
of the inserted fake sources were measured by manual photometry adopting the same parameters, i.e. same
aperture, used for the real sources. The simulations establish the relationship between source signal detected
in the photometric aperture, and total source signal actually collected in the detector. Signal in the
photometric aperture can therefore be scaled to total signal, but at this point the units are detector units, or
ADUgs. The relationship between source signal deposited in the detector and actual source flux-density in mJy is
determined by two further scaling factors.  (a) A filter specific ISOCAM calibration factor relating stabilised
source signal, in detector units, to mJy, i.e. 1 ADU per gain per second (ADUgs) $\equiv$ 0.43\,mJy for LW2 and
1 ADUgs $\equiv$ 0.51\,mJy for LW3 (Delaney \& Ott \cite{ISOCAM}); and (b) a correction for detector responsive
transient effects which cause the signal from faint sources to fall below that expected on the basis of bright
reference source measurements. The derivation of that scaling factor, in the context of the reduction algorithm
applied here, is described in Metcalfe et al. (\cite{2003A&A...407..791M}) and references therein.

\begin{table}[\columnwidth]     % TABLE 5
    \centering
\caption[]{References to ISOCAM sources in existing archives. The first column lists the source number from
Tables~\ref{lw3} and \ref{lw3_add}. The key to the references in Cols. 2 to 8 is Dressler \& Gunn
(\cite{1992ApJS...78....1D}, DG), Czoske et~al. (\cite{2001A&A...372..391C}, CKS), Schneider et~al.
(\cite{1986AJ.....92..523S}, SDG), Smail et~al. (\cite{1997ApJS..110..213S}, SDC), McLean \& Teplitz
(\cite{1996AJ....112.2500M}, MT), Butcher \& Oemler (\cite{1978ApJ...219...18B}, BO), Soucail et~al.
(\cite{2000A&A...355..433S}, S) and Pickles \& van der Kruit (\cite{1991A&AS...91....1P}, P). The table does not
include the four stars or the quasar PC 0023+1653.\label{other}}
   \begin{tabular}{cccccccl}
\hline \hline
id    & DG & CKS & SDG & SDC & MT  & BO  &Others\\
\hline \noalign{\smallskip}
05    &198 &262  &223  &     &     &123  &      \\
06    &278 &     &     &     &64   &     &      \\
08    &264 &267  &113  &     &22   &34   &S01   \\
09    &257 &282  &     &     &6    &     &      \\
10    &237 &289  &     &     &     &     &      \\
12    &    &381  &     &     &     &     &      \\
17    &12  &     &     &     &     &     &      \\
18    &47  & 444 &     &797  &4    &     &      \\
21    &119 &453  &     &     &     &     &      \\
22    &43  &459  &146  &834  &     &     &      \\
23    &23  &     &     &883  &     &     &      \\
24    &    &471  &     &     &     &     &      \\
27    &3   &474  &     &928  &     &     &      \\
1s    &195 &280  &     &     &50   &     &      \\
3s    &246 &304  &     &     &55   &     &      \\
4s    &101 &412  &186  &573  &11   &87   &P61   \\
 \hline
      \end{tabular}
\end{table}
\section{Results\label{sec:results}}          % SECTION 4

The merged 15\,$\mu$m map (Fig.~\ref{lw3_fig}) is overlaid on a V-band image of the cluster taken with the Very Large
Telescope (VLT).

The list of sources detected in the two individual 15\,$\mu$m maps is given in Table~\ref{lw3} and a
supplementary list is given in Table~\ref{lw3_add} (see Sect. \ref{sec:obsda}). The name of the ISOCAM source is
composed of the satellite acronym (ISO), the partial name of the cluster (Cl0024), and an identification number
assigned to each source.

The 7\,$\mu$m fluxes for the stars, and their [7\,$\mu$m]/[15\,$\mu$m] flux ratios, are given in
Table~\ref{lw2}. The labels of the 7\,$\mu$m sources are taken to be the same as their 15\,$\mu$m counterparts.
The fourth star detected at 15\,$\mu$m (ISO\_Cl0024\_01) is outside the boundary of the 7\,$\mu$m raster. No
other sources were detected above the 5 $\sigma$ limit of $\sim$ 400\,$\mu$Jy.

The column named ``Precision" in Tables~\ref{lw3} and \ref{lw3_add} reflects the repeatability of the
photometric results in the fake source simulations. For each source brightness the precision is the 1$\sigma$
scatter found in the recovered fluxes of fake sources of similar brightness inserted into the raw data. It
should be noted that the ratio signal/precision is not, however, the signal-to-noise ratio; nor is it the
significance of a source detection.  The precision, as defined here, includes, for example, signal deviations
caused by residual glitches around the map and by variations in residual background level across the map, which
are of course factors that can influence the accuracy of source photometric measurements. However, the
significance of an individual source detection must be determined through an examination of the pixel-to-pixel
signal variations in the map local to the source. All of the sources listed in Tables~\ref{lw3} and
\ref{lw3_add} have significance values greater than 5$\sigma$. The quoted precision is a measure of the relative
calibration accuracy within the sample recorded. The absolute flux calibration is, in addition, affected by
factors such as imperfections in the knowledge of the detector response and responsive transient correction, and
differences between the spectral shapes of measured sources with respect to the canonical calibration source for
ISOCAM, which has a stellar spectrum. Experience over many surveys suggests an absolute calibration accuracy of
$\pm$\,15\% for ISOCAM faint-source photometry due to accumulated systematic effects. The measured colour ratios
for the stars listed in Table~\ref{lw2} are consistent with this, and give some confidence in the absolute
calibration at the $\pm$\,15\% level.

The number of sources in the primary and supplementary lists are 30 and 5, respectively, yielding a total of 35
sources. A search was performed for counterparts of the ISOCAM sources at other wavelengths, using a search
radius of 6\arcsec~in HEASARC\footnote{http://heasarc.gsfc.nasa.gov/db-perl/W3Browse/w3browse.pl} and NASA/IPAC
Extragalactic Database\footnote{The NASA/IPAC Extragalactic Database (NED) is operated by the Jet Propulsion
Laboratory, California Institute of Technology, under contract with the National Aeronautics and Space
Administration.}. Of the 30 sources in the primary list (Table~\ref{lw3}), 19 are identified with catalogued
sources. Of these nineteen, four are stars, one is a quasar (PC 0023+1653), ten are cluster members, one is a
foreground galaxy and three are background galaxies. As for the 5 supplementary sources (Table~\ref{lw3_add}),
three are identified with cluster galaxies. There is a total of 13 sources with unknown redshift and all have
optical counterparts (Fig.~\ref{lw3_fig}).

The median flux-density of the unknown-redshift sources is about 250\,$\mu$Jy. The expected number of 15
\,$\mu$m sources lensed by the cluster down to that flux limit is about $20\pm10$, based on the log N - log S
distribution of Metcalfe et al. (\cite{2003A&A...407..791M}). So the expected number of background sources is
sufficient to account for the observed number of unknown-redshift sources.

\begin{figure*}             % FIGURE 3
\includegraphics[width=\textwidth,height=0.8\textwidth]{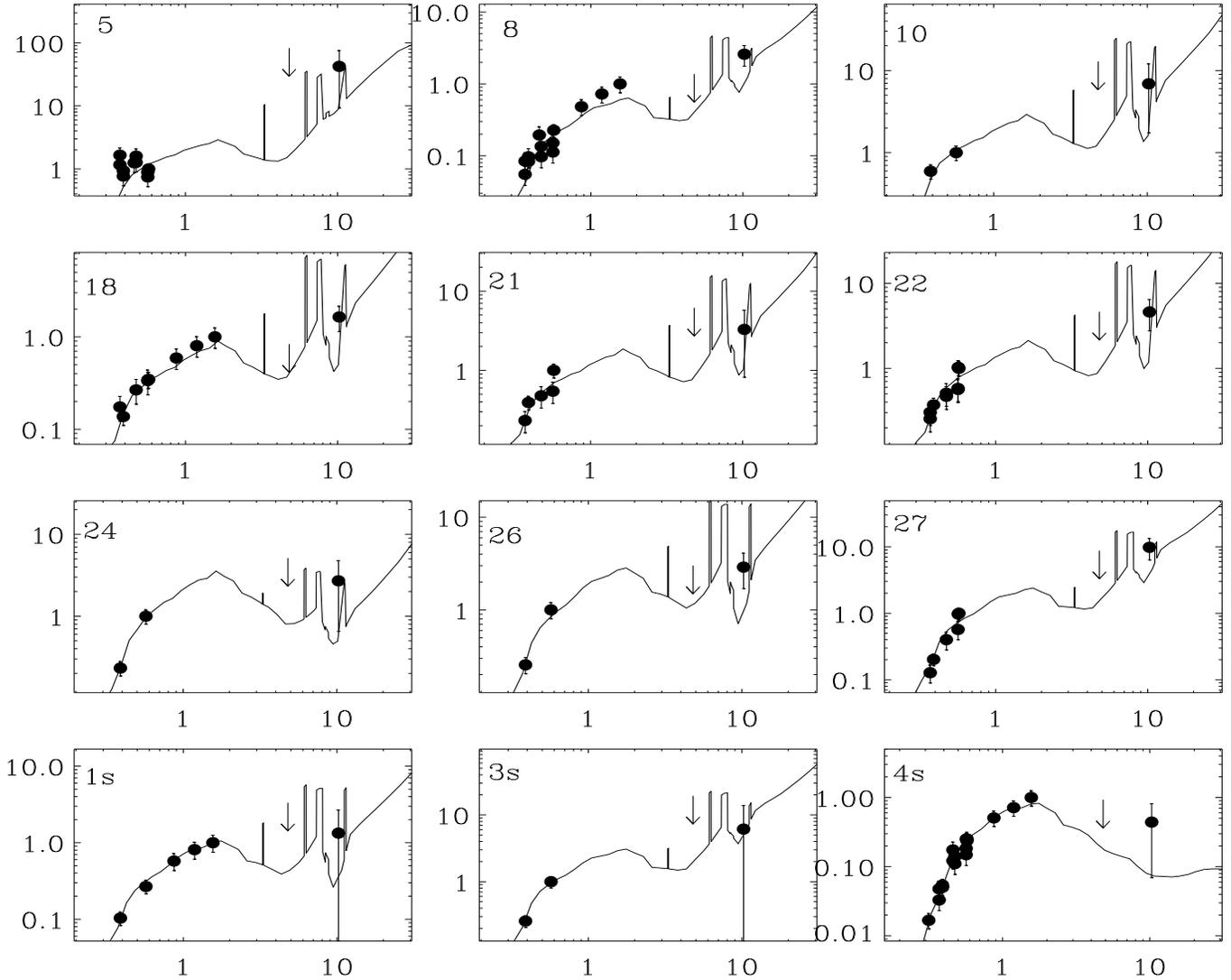}%}
\caption[]{SEDs for cluster members in Cl 0024+1654 detected at 15\,$\mu$m. The horizontal axis is the wavelength
in the cluster rest frame and the vertical axis is the flux density (in normalized units). The data points are
given by black dots and the model fit to the SED by a continuous line. \label{andrea}}
\end{figure*}

\subsection{Spectral Energy Distributions\label{sec:spectral} } % SECTION 4.1

Spectral energy distributions were computed for the cluster galaxies using the 15\,$\mu$m fluxes in
Tables~\ref{lw3} and \ref{lw3_add}, and a 5$\sigma$ upper limit of 400 $\mu$Jy for the 7\,$\mu$m fluxes. The
spectral range of the ISOCAM sources was extended by including measurements in the optical photometric bands and
the near-infrared. The values were retrieved from the NASA/IPAC Extragalactic Database. The references to
extensive observations of the ISOCAM sources from a wide range of catalogues are listed in Table~\ref{other}.

The SEDs were modelled using the program GRASIL (Silva et al. \cite{1998ApJ...509..103S}). These models have
already been used by Mann et al. (\cite{2002MNRAS.332..549M}) to fit the SEDs of galaxies detected by ISOCAM in
the {\em Hubble Deep Field South.} Given the limited photometric accuracy and, for some galaxies, the limited
amount of available photometric data, it is not possible to obtain an accurate model of the SED of each
individual galaxy. Instead, observed SEDs are compared with models representative of broad classes of spectral
types. The number of models to be considered needs to be adequate to reflect the quality of the data and the
questions to be answered (see, e.g., Rowan-Robinson \cite{2003MNRAS.344...13R}).  The redshifts of the galaxies
are known and hence it was possible to use more than the five models of Mann et al. (\cite{2002MNRAS.332..549M})
without the risk of multiple solutions.

\begin{table*}%[0.5\columnwidth]            %TABLE 6
    \centering
\caption[]{Results from model SEDs for known cluster galaxies in Cl 0024+1654. The columns list: LW3
identification number; best-fit SED model; number of data points in the observed SED; $\chi^2$ of the fit;
rejection probability of the fit; other acceptable models (rejection probability 95\% or less); 15\,$\mu$m
k-correction; luminosity at 15\,$\mu$m (h=0.7, $\Omega_m=0.3$, $\Omega_l=0.7$) using the transformation from
Boselli et al. (\cite{1998A&A...335...53B}), and k-corrected using best-fit SED; total infrared luminosity
including the k-correction using the best-fit SED and computed using the transformation from Elbaz et~al.
(\cite{2002A&A...384..848E}); SFR from total IR luminosity using the transformation of Kennicutt
(\cite{1998ARA&A..36..189K}); position in the V-I vs. I colour-magnitude diagram (Fig. \ref{fig:color}) where
`MS' stands for `Main Sequence', `BO' for `Butcher-Oemler', `U' for `Unknown'; EW\,[\ion{O}{ii}] line from
Czoske et~al. (\cite{2001A&A...372..391C}); SFR computed from the [\ion{O}{ii}] emission line using the
transformation to the optical SFR; ratio infrared to optical SFR. \label{sfrir}}
  \begin{center} \leavevmode\scriptsize\begin{tabular}{cccccccccccccc}
\hline
         \hline
id & Best-fit & $\sharp$ of & $\chi^2$&Rejection  & Other SED & k-corr.&L$_{15\mu \mathrm{m}}$& L$_\mathrm{IR}$&SFR[IR]&Colour&EW[\ion{O}{ii}]   & SFR[\ion{O}{ii}]& SFR[IR]/\\
   & model    & data points &         &Probability& models    &             & (L$_\odot$)          &  (L$_\odot$)   & (M$_\odot$\,yr$^{-1}$)& Fig. 8 &(\AA)&(M$_\odot$\,yr$^{-1}$)&   /SFR[\ion{O}{ii}]\\
\hline \noalign{\smallskip}
 05& S    & 12 & 17.8 & 96.2 &none              & 1.45 & 2.03e+09 & 9.56e+10 & 16 & U  & -    &  - &  -  \\
 08& SB+1 & 15 & 23.4 & 97.6 &none              & 1.80 & 9.66e+09 & 4.54e+11 & 77 & BO & -    &  - &  -  \\
 10& S    & 4  & 1.9  & 79.8 &S, SB+1    & 1.46 & 2.21e+09 & 1.04e+11 & 18 & BO & -    &  - &  -  \\
 18& S    & 11 & 3.9  & 13.4 &E, S, SB+1  & 1.46 & 7.85e+09 & 3.69e+11 & 63 & BO & -    &  - &  -  \\
 21& S    & 7  & 5.4  & 74.7 &SB+1, S         & 1.46 & 2.11e+09 & 9.93e+10 & 17 & BO & -    &  - &  -  \\
 22& S    & 11 & 7.6  & 52.6 &S, SB+1  & 1.46 & 3.96e+09 & 1.87e+11 & 32 & BO & 17.5 & 4.9& 6.5\\
 24& S    & 4  & 1.2  & 79.8 &E, SB+1, S       & 1.42 & 2.07e+09 & 9.74e+10 & 17 & MS & 5.0  & 0.8& 21.3\\
 26& S    & 4  & 1.4  & 79.8 &S, SB+1       & 1.73 & 4.51e+09 & 2.17e+11 & 36 & -  & -    & -  &  -  \\
 27& SB+1 & 8  & 6.7  & 87.8 &SB              & 1.80 & 5.50e+09 & 2.59e+11 & 44 & MS & -    & -  &  -  \\
 1s& S    & 10 & 8.6  & 71.4 &E, E, S, SB+1   & 1.73 & 1.96e+09 & 9.23e+10 & 16 & BO & 9.6  & 1.0& 16.0\\
 3s& SB+1 & 4  & 1.3  & 79.8 &S, SB+1, E  & 1.80 & 1.59e+09 & 7.48e+10 & 13 & MS & -    &  - &  -  \\
 4s& E    & 20 & 9.4  & 7.3  &E, S, SB+1 & 0.78 & 9.87e+08 & 4.66e+10 & 8  & MS & -    &  - &  -  \\
   \hline
      \end{tabular}
      \end{center}
\end{table*}

In this work, 20 models were considered. They were taken from either the public GRASIL library, or built by
running the publicly available GRASIL code, or kindly provided by L.~Silva (private comm.). The 20 models
reproduce the SEDs of several kinds of galaxies, from early-type, passively evolving ellipticals (labelled `E',
in the following), to spiral galaxies (labelled `S'), and starburst galaxies (similar to the local examples
Arp220 and M82), as seen either at the epoch of the starburst event, or 1 Gyr after (labelled `SB' or `SB+1',
respectively). At variance with Mann et al. (\cite{2002MNRAS.332..549M}) we avoid considering models older than
the age of the Universe at the cluster redshift. Specifically, we consider models for galaxies with a formation
redshift $z=1$ (this is the redshift when stars start forming in the GRASIL models), corresponding to an age at
the cluster redshift of $\sim 3$ Gyr, and models with a formation redshift $z=4$, corresponding to an age at the
cluster redshift of $\sim 8$ Gyr.

The best-fitting SED model was determined by $\chi^2$ minimization, leaving the normalization of the model free.
In principle, a given SED model cannot be freely rescaled, since the parameters in the GRASIL code depend on the
mass of the galaxy. However, a fine tuning of the values of the GRASIL parameters only makes sense if the
photometric uncertainties of the observed galaxy are small enough, and the SED is very well covered by
observations, which is not the case here. As a matter of fact, the model SEDs of two galaxies of very different
mass, such as M82 and NGC6090, could not be distinguished after rescaling, in the wavelength range covered by
our observations, and with the accuracy of our photometric data points.

The SEDs are plotted in Fig. \ref{andrea} and summarized in Table \ref{sfrir} for all cluster galaxies detected
with ISOCAM, except for ISO\_Cl0024\_12, for which there are not enough photometric data. All the fits are
acceptable at the 1\% confidence level, and all, except two, at the 5\% confidence level. S and SB+1 models
provide the best-fit to most SEDs, but the fits are, in general, not unique. Among the acceptable fits (at the
5\% confidence level) there are also SB and E models, but the latter always underestimate the mid-infrared fluxes. We
are unable to distinguish between models with different formation redshifts.

Overall, the results show that most ISOCAM cluster sources have SEDs typical of actively star-forming galaxies,
although not necessarily observed in an exceptional starbursting phase.

\begin{figure}          % FIGURE 4
%\resizebox{\columnwidth}{!}{\includegraphics{ave_sed.eps}}
\resizebox{\columnwidth}{!}{\includegraphics{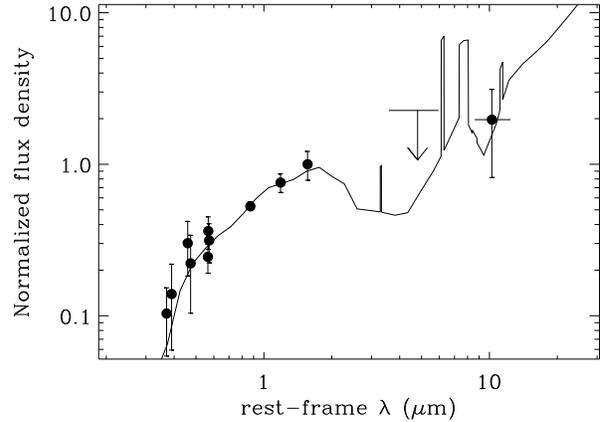}}  \caption{The average SED of 9 ISOCAM cluster
sources (the sources from the supplementary list have not been considered). Error bars give the rms values for
the flux density. An upper limit of 400 $\mu$Jy was used for LW2 at 7\,$\mu$m. The solid line shows a model fit
(SB+1) to the average SED.}\label{fig:ave_sed}
\end{figure}

The SEDs of the nine cluster sources from the primary list were combined assuming the mean redshift of the
ISOCAM cluster sources, and normalising the individual SEDs with the flux density in the rest-frame $H$ band,
either observed or as predicted by the best-fit SED models (see also Biviano et al. \cite{biviano04}).  We did
not consider the three supplementary sources; since their 15\,$\mu$m flux values are rather uncertain, including
these sources would increase the scatter of the average 15\,$\mu$m flux density value.  The resulting average
observed SED is shown in Fig. \ref{fig:ave_sed}, along with a SB+1 model fit to guide the eye (S models fit the
average SED equally well). The average SED provides a synthetic view of the spectral shape of the 15\,$\mu$m
cluster galaxies, but we refrain from a physical interpretation of it, which would require a detailed analysis
of the systematics inherent to its construction.

\subsection{Infrared and optical star formation rates for cluster members\label{sec:sfrmembers} } % SECTION 4.2

The mid-infrared emission, free from dust extinction, is a reliable tracer of star formation
(Genzel \& Cesarsky \cite{2000ARA&A..38..761G}). The infrared emission from a galaxy is the sum of various
contributions including:
\begin{description}
\item[1)] continuum emission from dust particles \item[2)] line emission from carriers of Unidentified Infrared
Bands (UIBs) \item[3)] line emission from ionized interstellar gas \item[4)] emission from the evolved stellar
population that dominates early-type galaxies \item[5)] non-thermal emission from radio sources.
\end{description}
The infrared spectrum of a galaxy depends on its morphological type and evolutionary status. For elliptical
galaxies, the spectrum is similar to a blackbody continuum at a temperature of 4000-6000 K with a
[7\,$\mu$m]/[15\,$\mu$m] flux ratio of about 4.5, while for spiral galaxies the ratio of the
[7\,$\mu$m]/[15\,$\mu$m] flux is around 1 (Boselli et al. \cite{1998A&A...335...53B}). Starburst galaxies are
characterized by a rapid increase in emission towards 15\,$\mu$m because of the contribution from very small
grains, which are dust particles with radii of $\sim$10\,nm that are abundant in star forming regions (Laurent
et al. \cite{2000A&A...359..887L}).

\begin{figure}[!t]           % FIGURE 5
%\resizebox{\hsize}{!}{\includegraphics{histo_lir.eps}}
\resizebox{\hsize}{!}{\includegraphics{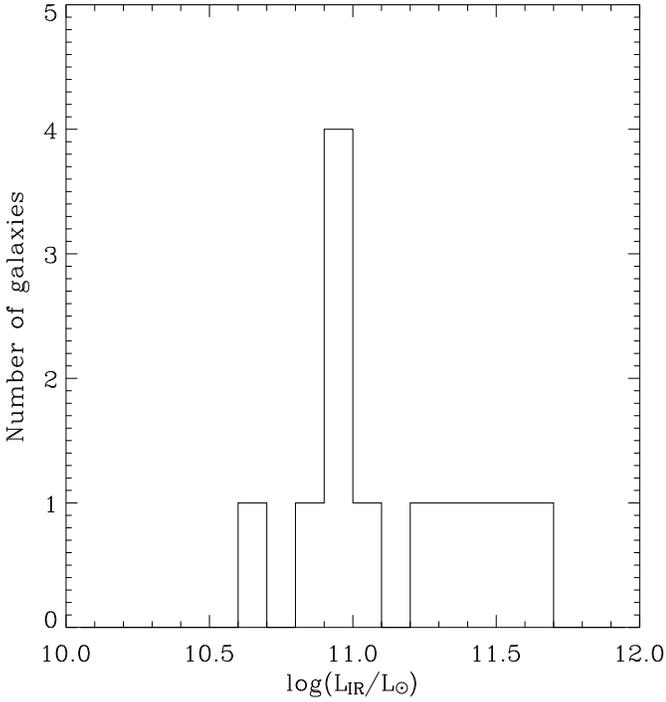}} \caption[]{The total infrared luminosity distribution for
cluster members. The median value for L$_\mathrm{IR}$ is $\sim1.0\times10^{11}$
$\mathrm{L}_\odot$}.\label{fig:lir_dist}
\end{figure}
The luminosities at 15\,$\mu$m, in units of solar luminosity, were obtained using the relationship (Boselli et
al. \cite{1998A&A...335...53B}):
\begin{equation}\label{eq:boselli}      %       EQUATION 1
\mathrm{L}_{15\,\mu m}=4\ \pi \ \mathrm{D_\mathrm{L}}^2\ \mathrm{F}_{15\,\mu \mathrm{m}}\ \delta_{15\,\mu
\mathrm{m}}
\end{equation}
where F$_{15\,\mu \mathrm{m}}$ is the flux at 15\,$\mu$m in mJy and
$\delta_{15\,\mu\mathrm{m}}=5.04\times10^{12}$ $\mathrm{Hz}$ is the bandwidth of the LW3 filter. The total
($8-1000$ $\mu$m) infrared luminosity was obtained from the empirical relation of Elbaz et al.
(\cite{2002A&A...384..848E}):
\begin{equation}\label{eq:elbaz} %       EQUATION 2
\mathrm{L}_\mathrm{IR}=11.1^{+5.5}_{-3.7}\times(\nu \ \mathrm{L}_{\nu}[15\mu \mathrm{m}])^{0.998}
\end{equation}
The total IR emission was also obtained directly from the best-fit SED model and provides an alternative
determination of the total infrared luminosity. The total IR luminosities obtained from the two methods were
compared and found to agree to within 32\%. We adopt the IR luminosities obtained from Equation~\ref{eq:elbaz}
for the cluster galaxies. The luminosities are listed in Table \ref{sfrir} and include the k-corrections
obtained from the best-fit SEDs. Depending on the best-fit model, the k-correction at the cluster redshift
ranges from 0.8 to 1.8 with a median (mean) of 1.5 (1.5). The distribution of the total infrared luminosities
for cluster galaxies is given in Fig.~\ref{fig:lir_dist}. The mid-infrared cluster members have total IR
luminosities between $4.7\times10^{10}$ $\mathrm{L}_\odot$ and $4.5\times10^{11}$ $\mathrm{L}_\odot$, with a median value of
$1.0\times10^{11}$ $\mathrm{L}_\odot$. Six of the 12 galaxies have total infrared luminosities above $10^{11}$ $\mathrm{L}_\odot$,
which classify them as LIRGs ($10^{11}\ \mathrm{L}_\odot\leq \mathrm{L_{IR}} \leq 10^{12}\ \mathrm{L}_\odot$, e.g. Genzel \&
Cesarsky \cite{2000ARA&A..38..761G}), and four more have infrared luminosities above $9\times10^{10}$
$\mathrm{L}_\odot$. The star formation rates in units of solar masses per year, SFR[IR], were derived using the relation of
(Kennicutt \cite{1998ARA&A..36..189K}):
\begin{equation}\label{eq:kennicutt}  %       EQUATION 3
\centering
 SFR[IR] \simeq 1.71 \times 10^{-10}\ (\mathrm{L_{IR}} /\mathrm{L}_\odot)
\end{equation}
The values of SFR[IR] are listed in Table~\ref{sfrir}. The infrared SFRs range between $\simeq 8$
M$_\odot$\,yr$^{-1}$ and $\simeq 77$ M$_\odot$\,yr$^{-1}$, with a median (mean) value of 18 (30) $\mathrm{M}_\odot\,\mathrm{yr}^{-1}$.

Half of the 15\,$\mu$m cluster galaxies are LIRGs and four more are within 1$\sigma$ of the luminosity of a
LIRG. Star forming episodes that are enshrouded by dust are commonly found in these galaxies in the field. The
cause of the starburst phase of LIRGs in the local universe is still debated, but in the range of infrared
luminosities measured with ISOCAM the starburst seems to be linked to mergers or interactions between pairs of
spiral galaxies of unequal size and single galaxies in about one-quarter of the sources (Hwang et al.
\cite{1999ApJ...511L..17H}; Ishida \& Sanders \cite{2001AAS...198.3416I}). The cause of the starburst in the
15\,$\mu$m sources of Cl 0024+1654 seems to be consistent with what is observed in the local universe because of
the evidence for interactions and mergers in the HST maps (see Sect. \ref{sec:color}).

\begin{figure*}[p\textwidth]    % FIGURES 6 & 7
\centering
%\mbox{\subfigure[]{\includegraphics*[width=\columnwidth,height=\columnwidth]{ak_xy_a.eps}\label{fig:kernel}}
%\quad
%\subfigure[]{\includegraphics*[width=\columnwidth,height=\columnwidth]{cumdist_a_in.eps}\label{fig:radial}}}
\mbox{\subfigure[]{\includegraphics*[width=\columnwidth,height=\columnwidth]{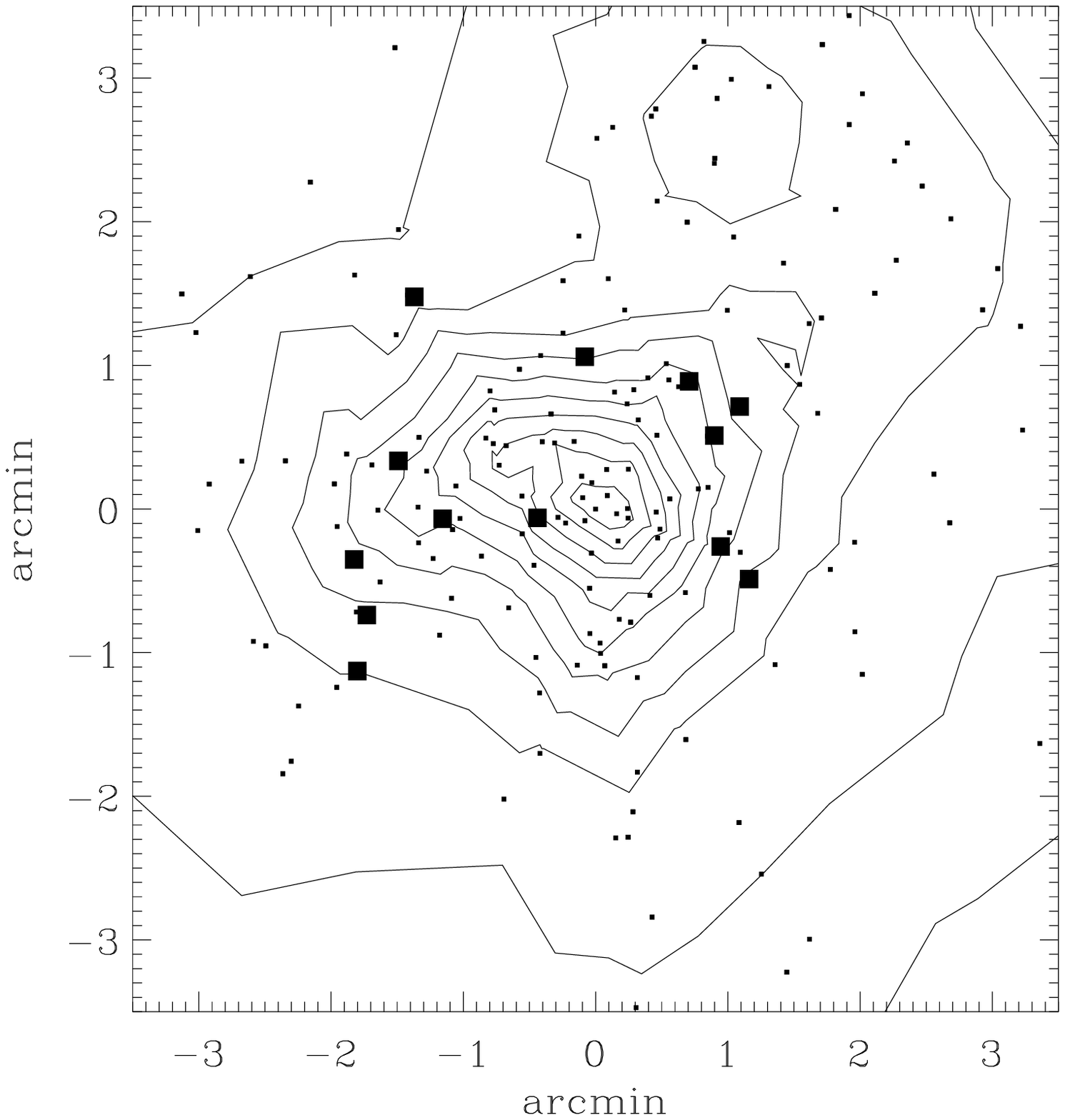}\label{fig:kernel}}
\quad \subfigure[]{\includegraphics*[width=\columnwidth,height=\columnwidth]{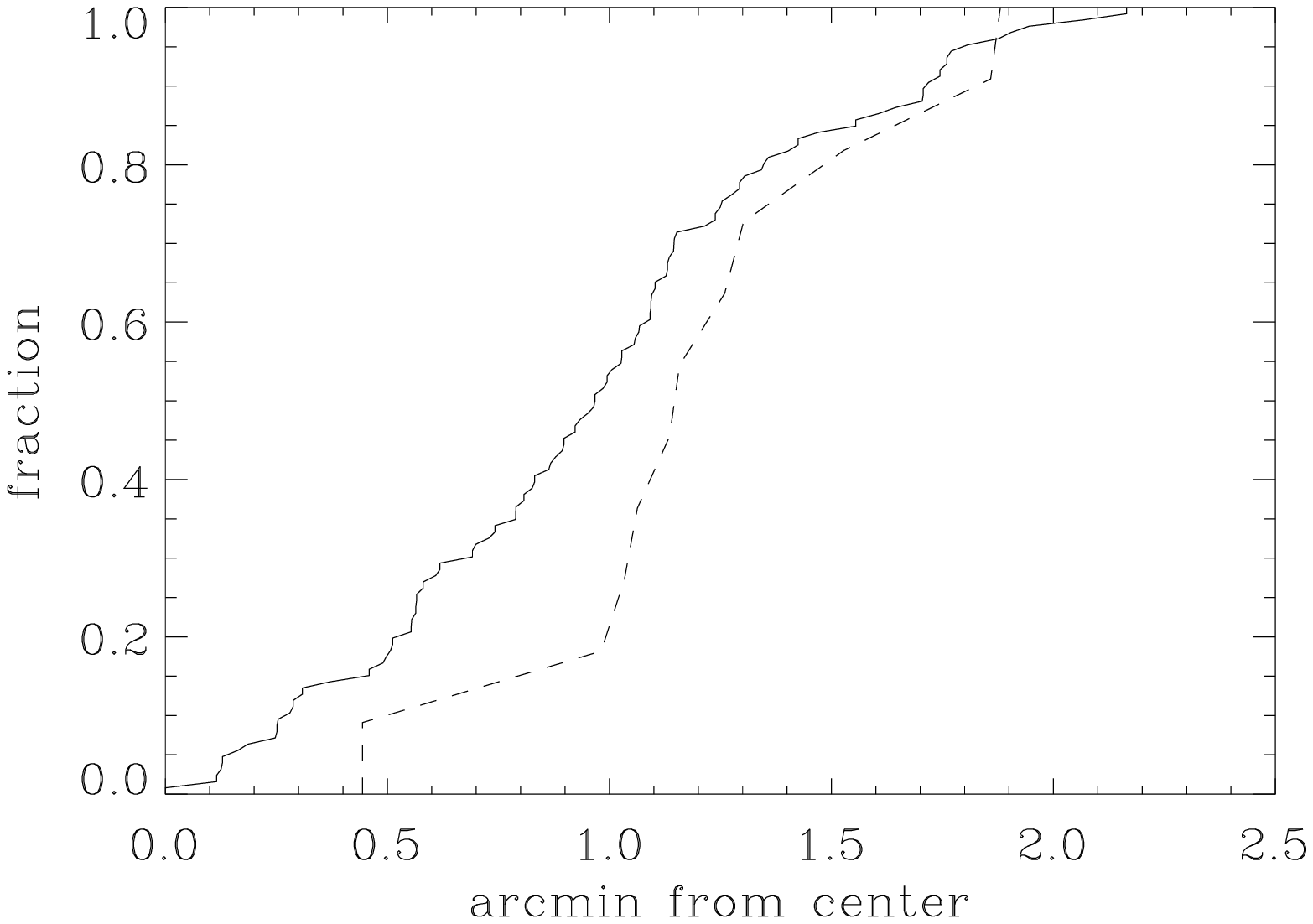}\label{fig:radial}}}
\caption{(a) Adaptive Kernel map for cluster members. Black boxes represent the 15\,$\mu$m sources while dots
and isodensity contours trace the distribution of component-A galaxies. (b) Radial distribution from the centre
of component-A galaxies (continuous line) and 15\,$\mu$m cluster galaxies (dashed line) within the region
covered by the ISOCAM observations. \label{fig:kernel_map}}
\end{figure*}
The [\ion{O}{ii}] line is often used to determine the SFR in galaxies at intermediate redshift, when the Balmer
hydrogen lines are outside the spectral range (e.g. Barbaro \& Poggianti \cite{1997A&A...324..490B}; Jansen et
al. \cite{2000ApJS..126..331J}). There are several problems associated with this line. [\ion{O}{ii}] is subject to dust
extinction, and its strength depends heavily on the metallicity and ionization of the interstellar medium
(Kennicutt \cite{1998ARA&A..36..189K}; Jansen et al. \cite{2000ApJS..126..331J}). Furthermore, Balmer lines are
more correlated to UV emission from young stars than [\ion{O}{ii}]. Therefore, the uncertainties in SFR computed from
[\ion{O}{ii}] are very large. Nevertheless, there may be times when it is the only indicator available, in which case
finding ways to correct the determination for dust extinction will be important.  It may well yield only a lower
limit on the SFR.

The Equivalent Width (EW) of the [\ion{O}{ii}] line was available from Czoske et al. (\cite{2001A&A...372..391C}) for
only three of the mid-infrared cluster members. The [\ion{O}{ii}] line is redshifted into the V band and can be used to
compute the optical SFR provided that the V-band magnitude is known. The luminosity L$_\mathrm{v}$ was derived
from the V-band magnitude in units of erg\ s$^{-1}$\ $\AA^{-1}$ using:
\begin{equation}         %       EQUATION 4
\mathrm{L_V} = 4\ \pi \
\mathrm{D}_\mathrm{L}^2\times(3.08\times10^{24})^2\times10^{-0.4\mathrm{m_v}}\times3.92\times10^{-9}
\end{equation} where $\mathrm{D}_\mathrm{L}$ is in Mpc and m$_\mathrm{v}$ is the apparent magnitude in the V-band. The
luminosity of the [\ion{O}{ii}] line is given by L[\ion{O}{ii}] = EW[\ion{O}{ii}]$\times$ L$_\mathrm{V}$ in
units of erg s$^{-1}$. Finally, the SFR was obtained using SFR[\ion{O}{ii}] $\sim 1.4\times10^{-41}$
L[\ion{O}{ii}], in units of $\mathrm{M}_\odot\,\mathrm{yr}^{-1}$ (Kennicutt \cite{1998ARA&A..36..189K}). The
SFRs derived from the EWs of the [\ion{O}{ii}] lines are listed in Table~\ref{sfrir} and should be regarded as
lower limits because no correction for extinction was applied. The mean value of the optical SFR is 2.2
$\mathrm{M}_\odot\,\mathrm{yr}^{-1}$ and increases to 3 $\mathrm{M}_\odot\,\mathrm{yr}^{-1}$ when the canonical
value of 1 mag at H$\alpha$ (Kennicutt \cite{1992ApJ...388..310K}), for extinction in the optical, is applied.
The values of SFR[\ion{O}{ii}] show a wide range, and are sometimes more than 1 order of magnitude lower than
those obtained in the infrared. Recent data show that there is a better agreement when the SFRs are computed
from the H$\alpha$ line and the mid-infrared (Kodama et al. \cite{astro-ph/0408037}). The ratios of
SFR[IR]/SFR[\ion{O}{ii}] range from 7 to 21 with a median value of 16 (section \ref{sec:sfrmembers}). Therefore
the vast bulk of the star formation is missed when the [\ion{O}{ii}] line emission is used. A large fraction of
the star formation in Cl 0024+1654 is enshrouded by dust (see Cardiel et al. \cite{cardiel} for a thorough
discussion on the comparison of SFR-estimators). Duc et al. (\cite{2002A&A...382...60D}) and Balogh et al
(\cite{2002MNRAS.335...10B}) have arrived at a very similar conclusion for the cluster Abell 1689, which has
many mid-infrared sources. Similar results have also been obtained for field galaxy samples (see, e.g., Charlot
et al. \cite{charlot}).

In our sample, AGN activity does not seem to be a major contributor to the energy output of the systems.
Measurements of the H$\alpha$ and [\ion{N}{ii}] emission lines are required to further determine the AGN contribution to
the mid-infrared emission (Balogh et al. \cite{2002MNRAS.335...10B}) and future radio observations could also
help in determining the AGN contribution (Rengarajan et al. \cite{1997MNRAS.290....1R}; Dwarakanath \& Owen
\cite{1999AJ....118..625D}).

\section{Spatial, dynamical and colour properties of the 15\,$\mu$m cluster galaxies\label{sec:distribution}} % SEC 5

\subsection{Spatial and velocity distributions}  % SECTION 5.1

The spatial and radial distributions of the cluster galaxies, with and without detectable 15\,$\mu$m emission,
are given in Fig.~\ref{fig:kernel_map}. The isodensity contours and dots in Fig.~\ref{fig:kernel} refer to
galaxies belonging to component A, while the black boxes represent the cluster members with 15\,$\mu$m emission.
The figure shows that the cluster members detected at 15\,$\mu$m are less concentrated than the cluster galaxies
not detected in the mid-infrared. The Kolmogorov-Smirnov test reveals that there is only a 4\% probability that
the two distributions are drawn from the same parent one.

The velocity distributions for cluster galaxies with and without detectable 15\,$\mu$m emission are given in
Fig.~\ref{fig:vel_disp}. The counterparts of the 15\,$\mu$m sources have redshifts which place them in Component
A - the larger of the two interacting components. One source, ISO\_Cl0024\_12, with $z = 0.386$ is close to the
boundary. The velocity dispersion computed in the cluster rest-frame (Fig~\ref{fig:vel_disp}), using the
biweight estimator (Beers et al. \cite{1990AJ....100...32B}) in the central region of the ISO map, is
$\sigma_\mathrm{v}=914^{+59}_{-56}$ km s$^{-1}$ for a sample of galaxies, while it is
$\sigma_\mathrm{v}=979^{+238}_{-193}$ km s$^{-1}$ for the 15\,$\mu$m galaxies. The 15\,$\mu$m sources are
associated mainly with spiral and emission-line galaxies, and studies of nearby clusters indicate a larger
velocity distribution for this population relative to the general cluster population %from the same population
(Biviano et al. \cite{1997A&A...321...84B}).

\begin{figure*}%[!h]                %  FIGURE 7
\centering \resizebox{1.3\columnwidth}{!}{
\includegraphics[angle = 90]{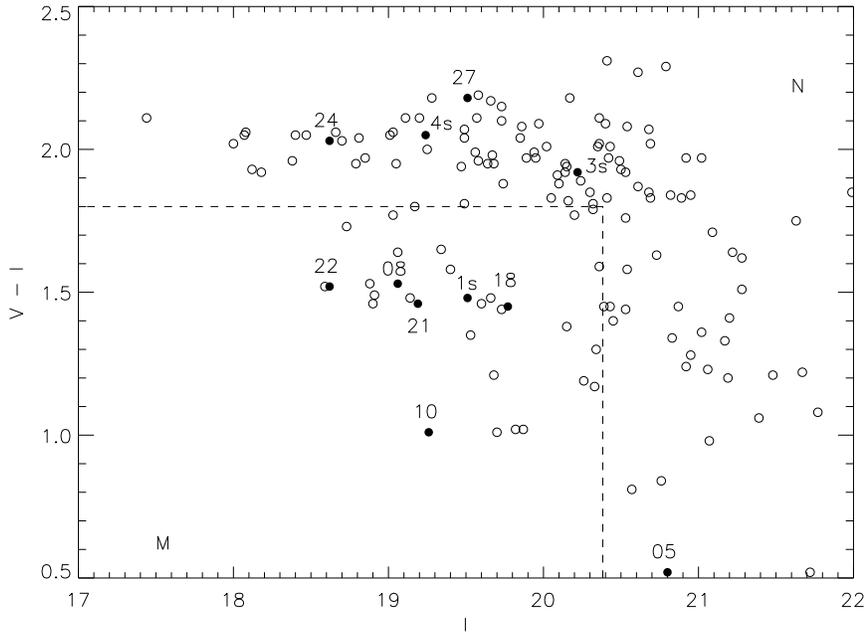}}
\caption[]{V - I colour-magnitude diagramme for component-A galaxies (empty circles) within the region mapped by
ISOCAM. The eleven 15\,$\mu$m cluster galaxies (filled circles) are identified by their ISO source number from
Tables \ref{lw3} and \ref{lw3_add}. The V and I magnitudes were not available for source ISO\_Cl0024\_12. The
galaxies in section M, which is defined by V - I $<$ 1.8 and I $<$ 20.4, are BO galaxies. Surprisingly, five of
the 15\,$\mu$m sources are in section N and not associated with BO galaxies. \label{fig:color}}
\end{figure*}

\begin{figure*}[]                   %  FIGURE 8
\centering
\begin{minipage}{\textwidth}
\resizebox{\hsize}{!}{
%\includegraphics*[width=6cm]{icon_06.eps}
%\includegraphics*{icon_17.eps}
%\includegraphics*{icon_14.eps}
%\includegraphics*{icon_29.eps}
%\includegraphics*{icon_36.eps}
%\includegraphics*{icon_42.eps}}
%\rotatebox{-90}{\includegraphics*{cluster_imagettes_mod.ps}}}
%\includegraphics*{cluster_imagettes.ps}}
{\includegraphics*{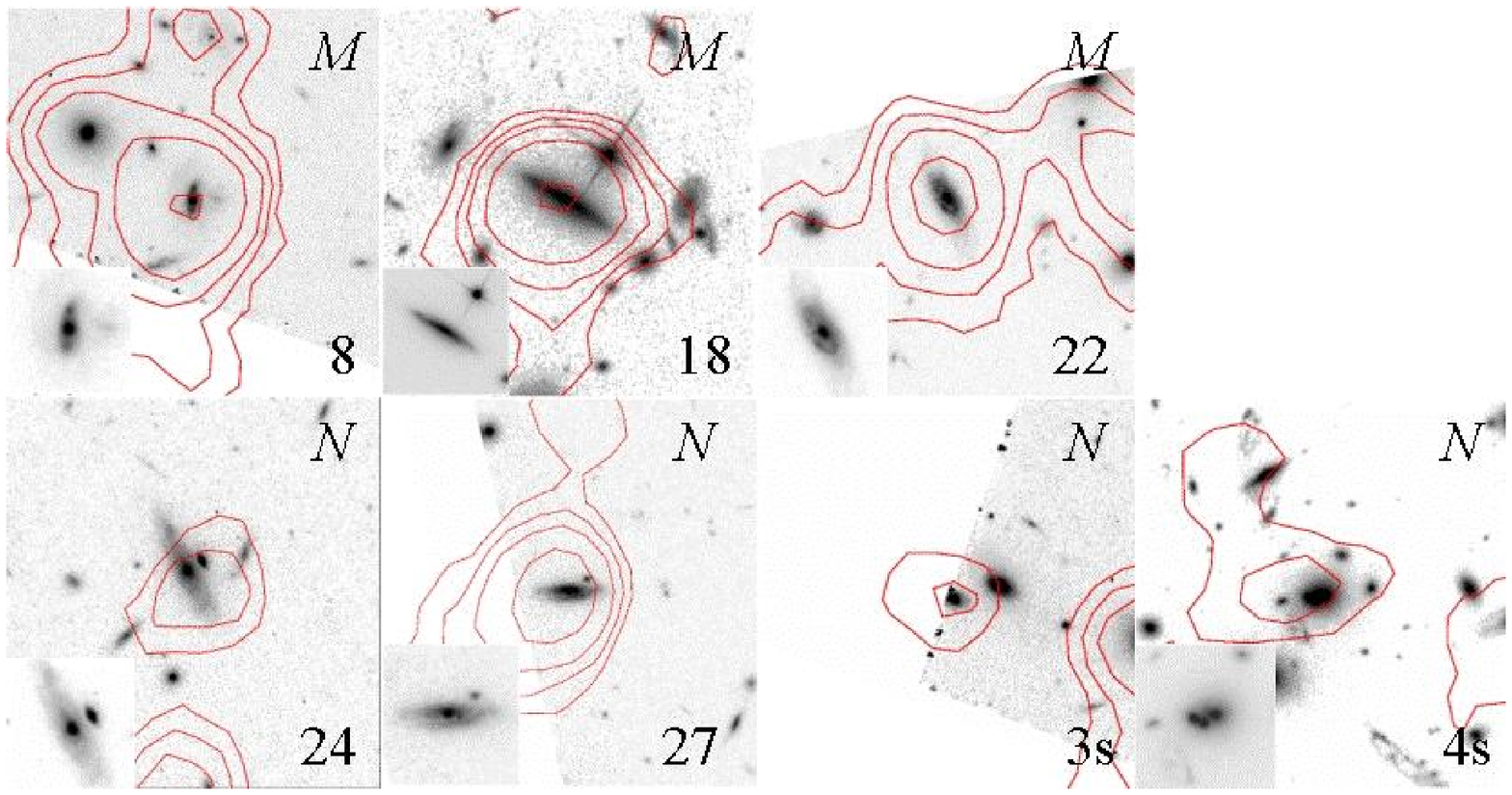}} }\caption[]{Miniature maps obtained from HST. The maps are centred on the
15\,$\mu$m source coordinates and labelled with the source number in Tables~\ref{lw3} and \ref{lw3_add}. The
letters M and N at the top right of the images give the positions of the cluster sources in the colour-magnitude
diagramme (Fig.~\ref{fig:color}). All images are $30\times30$ arcsec$^2$. Insets show details of the optical
counterpart. North is up, East to the left. \label{hst2}}
\end{minipage}
\end{figure*}
\subsection{Colour-magnitude diagramme\label{sec:color}}% SECTION 5.2

The colour-magnitude diagramme for cluster galaxies is given in Fig.~\ref{fig:color}. It provides an important
link between the optical and mid-infrared properties of the ISOCAM galaxies. Apparent magnitudes in the V and I
bands are available from Czoske et al. (\cite{2001A&A...372..391C}) for 11 of the 13 ISOCAM cluster galaxies.
The circles represent component-A galaxies within the region mapped by ISOCAM. The spectroscopically confirmed
mid-infrared cluster galaxies are indicated by filled circles and are labelled with the ISO source number.
Section M of Fig.~\ref{fig:color} contains the galaxies that satisfy the BO definition, given in section
\ref{sec:sec1}. Six of the 15\,$\mu$m galaxies fall in section M and satisfy the BO requirement. Only 20\% of the
cluster galaxies in section M are detected at 15\,$\mu$m. However, this fraction increases to $\sim$38\% when
only galaxies brighter than $\mathrm{m_I}= 19.8$ are considered. The non-detections of fainter I-band sources
are probably due to the sources falling below the sensitivity limits of ISOCAM.

All 15\,$\mu$m cluster sources have in common an excess of mid-infrared emission that sets them apart
from the rest of the cluster members. In at least some of these systems, one would expect that the star forming
activity is primarily triggered by interactions and mergers. Recent changes in the galaxy population reveal
themselves in the mid-infrared and help identify the processes that cause the burst of obscured star formation.
The infall of smaller groups into the cluster environment provides a way of promoting slow encounters and
mergers within clusters (Mihos \cite{2004cgpc.symp..278M}). Slow encounters between galaxies are better able to
drive instabilities than fast encounters. Complex interactions of the kind that may occur between galaxies in
this cluster have been studied in the local Universe in Hickson Compact Groups (see, e.g., L{\' o}pez-S{\'
a}nchez et al. \cite{2004ApJS..153..243L} and references therein).

We looked for indications of ongoing and past interaction events among the 15\,$\mu$m cluster sources, either in
their morphologies, or in their spectra. A HST image was available for part of Cl 0024+1654 from the MORPHS
collaboration (Smail et al. \cite{1997ApJS..110..213S}). This image is limited to the South-Eastern region of
the cluster core and contains seven of the 15\,$\mu$m cluster sources. For these galaxies, miniature maps
centred on the ISOCAM source coordinates and having an area of $30\times30$ arcsec$^2$ were obtained. The
miniature maps are presented in Fig.~\ref{hst2}.

Miniature maps are available for 3 of the BO galaxies in section M of Fig.~\ref{fig:color}. The HST maps show
that ISO\_Cl0024\_18 and ISO\_Cl0024\_22 do not have companions within a radius of 5\arcsec\ while
ISO\_Cl0024\_08 appears to be interacting with at least another galaxy.

HST miniature maps for four of the non-BO 15\,$\mu$m emitters in section N are also given in Fig.~\ref{hst2} and
each has at least one nearby companion. ISO\_Cl0024\_4s appears to be merging because it has multiple nuclei and
has at least two other companions. ISO\_Cl0024\_24 and ISO\_Cl0024\_3s show evidence of ongoing interactions
(e.g. tidal bridges) with at least one nearby galaxy.

Even though the number of infrared sources with HST maps is small, there seems to be a difference between the BO
and non-BO galaxies in the sense that the former tend not to have nearby companions whereas the latter do. The
optical counterparts of the 15\,$\mu$m sources on the main colour-magnitude sequence are not passive early type
galaxies as one would expect from their position in the colour-magnitude diagramme. Rather, the optical
counterparts appear to be galaxies that are involved in interactions and mergers in the HST images.

There is indirect support for this apparent difference between the BO and non-BO galaxies from the results of
the SED fitting process.  Five of the six 15\,$\mu$m galaxies in the BO region of the diagram have SEDs that are
best-fit by S-type models.  On the other hand, two out of four 15\,$\mu$m galaxies on the main sequence of the
colour-magnitude diagram have SEDs that are best-fit by SB+1-type.  If the starburst events are triggered by
interactions, there is marginal evidence from the SED analysis that the 15\,$\mu$m galaxies on the main
colour-magnitude sequence have suffered more interaction events in the recent past than the BO 15\,$\mu$m
galaxies.

Given the scarce morphological data, and the fact that the SED fitting results are not very well
constrained (SB+1-type models also provide acceptable fits to the SEDs of BO galaxies, see Table~\ref{sfrir}),
we consider the evidence for a difference in the interaction properties of BO and non-BO 15\,$\mu$m galaxies only
tentative. We conclude from our data that interaction events do trigger the star formation activity (and
stimulate the IR emission) in at least some 15\,$\mu$m cluster galaxies.

Obscuration by dust could explain why some 15\,$\mu$m sources lie on the cluster main sequence. We tested this
possibility by computing the de-reddened V and I magnitudes of the 15\,$\mu$m galaxies, using the dust-free
best-fit model SEDs, as given by the GRASIL code. As expected, once de-reddened the 15\,$\mu$m galaxies on the
colour-magnitude sequence move in the colour-magnitude diagramme toward the BO region, and in one case
(ISO\_Cl0024\_3s), it even becomes part of the BO category (Fig.~\ref{fig:color}). Dust obscuration at least
partially explains the unusual colours of the 15\,$\mu$m sources on the cluster main sequence.

\begin{table*}[ht2\columnwidth]             % TABLE 7
\caption[]{Summary of ISOCAM observations and results at 15\,$\mu$m for five clusters of galaxies.  The data was
obtained from Metcalfe et al. (\cite{2003A&A...407..791M}) for Abell 370, Abell 2218 and Abell 2390, Fadda et
al. (\cite{2000A&A...361..827F}) and Duc et al. (\cite{2002A&A...382...60D}) for Abell 1689, and this paper for
Cl 0024+1654.  The content of the columns in the table are as follows: name and redshift of the cluster, total
area scanned, sensitivity reported at the $5\sigma$ level, flux of the weakest reported source in $\mu$Jy, total
observation time, total number of sources detected including sources without redshifts and stars. Then number of
cluster galaxies, virial radius, virial mass, number of sources with $\mathrm{L_{IR}>9\times10^{10}\,L_{\sun}}$
detected and expected. The expected number of sources was obtained by comparison with Cl 0024+1654 as described
in the text. Virial radii and masses are from Girardi \& Mezzetti (\cite{2001ApJ...548...79G}) and King et al.
(\cite{2002A&A...383..118K}). \label{comp}}
\begin{center}
\leavevmode \small
      \begin{tabular}{cccccccccccc}
\hline
         \hline
\noalign{\smallskip}
Cluster& z    & Area        & Sensitivity& Faintest source&Obs. t. & Tot. n.    & Cluster & $\mathrm{R}_\mathrm{vir}$&$\mathrm{M}_\mathrm{vir}$& \multicolumn{2}{c}{n. of IR sources} \\
       &      &$(\arcmin^2)$&(5$\sigma$) &($\mu$Jy)       & (sec)  & sources    & galaxies&  ($\mathrm{h}^{-1}$ Mpc)                & ($\mathrm{h}^{-1}$ $10^{14}$ $\mathrm{M}_\odot$)         & Detected&Expected\\
\hline \noalign{\smallskip}
Cl0024 & 0.39 & 37.8        &   140      &  141           & 22615  & 35       & 13 & 0.94 &6.42 &10&  - \\
\noalign{\smallskip}
A370   & 0.37 & 40.5        &  350       &  208           & 22688  & 20       & 1  & 0.91 &5.53 &1&8 \\%D = 1981.7
\noalign{\smallskip}
A1689  & 0.18 & 36.0        & 450        & 320            & 9500   & 18       & 11 & 1.1  &5.7  &0&1   \\
\noalign{\smallskip}
A2390  & 0.23 & 7.0         & 100        & 54             & 29300  & 28       & 4  & 1.62 &20.35&0& 1\\
\noalign{\smallskip}
A2218  & 0.18 & 20.5        &  125       & 90             & 22000  & 46       & 6  & 1.63 & 18.27 & 0 &1 \\
%\noalign{\smallskip}
%A2219  & 0.23 & 13.4        &            & 248            & 4070   & 10       & 3  & 3.1  & 28.0 &2  &0\\%D=1138
%\noalign{\smallskip}
%A2219  & 0.23 & 13.4       &            & 248            & 4070   & 10       & 3  & 3.1  & 28.0 &2  &\\
\noalign{\smallskip}
       \hline
      \end{tabular}
\end{center}
\end{table*}
\section{Comparison of Cl 0024+1654 with other clusters observed by ISO\label{sec:compa}} %SECTION 6

Observations of clusters of galaxies at high redshift ($z>0.4$) with SCUBA have yielded an excess of
sub-millimeter sources with high values of the SFR (Best \cite{2002MNRAS.336.1293B}). Here we compare the
results of the observations of Cl 0024+1654 with four clusters of galaxies that were observed with ISOCAM at
15\,$\mu$m (Fadda et al. \cite{2000A&A...361..827F}; Metcalfe et al. \cite{2003A&A...407..791M}) to determine if
Cl0024+1654 has more LIRGs than the other clusters.  We have not included the cluster J 1888.16 Cl in Table
\ref{comp} because the relevant data have not been published (Duc et al. \cite{astro-ph/0404183}). However this
cluster seems to be comparable to Cl0024+1654 because it has many infrared sources and at least 6 of them are
confirmed LIRGs. The characteristics of the observations, and results for the five clusters, are summarized in
Table~\ref{comp}. The observed areas and sensitivities are comparable for the first three clusters in
Table~\ref{comp}, but the area is somewhat smaller for Abell 2218, and much smaller for the ultra-deep
observations of Abell 2390.

The number of 15\,$\mu$m sources that are identified with cluster galaxies ranges from 13 in Cl 0024+1654 to 1 in
Abell 370.  The number of 15\,$\mu$m sources that have fluxes consistent with LIRGs, within the precision of the
measurements is listed in Table \ref{comp}. In Cl 0024+1654 the number of LIRGs is 10 (Table \ref{comp}) and
includes 6 sources with L$_{IR} \geq 10^{11}$ $\mathrm{L}_\odot$ and 4 more with L$_{IR}$ between $9 \times 10^{10}$
$\mathrm{L}_\odot$ and $1\times10^{11}$ $\mathrm{L}_\odot$ (Table \ref{sfrir}).  The number of LIRGs detected in the other clusters is
either 0 or 1 (Table \ref{comp}). The virial mass and radius of Cl 0024+1654, Abell 370 and Abell 1689 are the
same to within 20\% and are considerably smaller than the values for the more massive clusters Abell 2218 and
Abell 2390.

However the ratio of the virial mass to area within the virial radius varies by less than 50\% for the 5
clusters.  We now compare the number of LIRGs in Cl 0024+1654 with those in the other clusters.  The comparison
is limited to LIRGs because it eliminates selection effects for the 15\,$\mu$m sources.  In any case most of the
15\,$\mu$m sources in the nearer clusters are too faint to be detected in the more distant clusters Cl 0024+1654
and Abell 370.  In the comparison, the number of LIRGs in Cl 0024+1654 was multiplied by the ratios of :
\begin{description}
\item[a)] the virial mass per unit area (since cluster mass is approximately proportioned to richness - see
Bahcall \& Cen \cite{1993ApJ...407L..49B}) of the cluster to that of Cl 0024+1654 \item[b)] the square
of the distance to the cluster and that to Cl 0024+1654 \item[c)] the observed solid angle of
the cluster to that of Cl 0024+1654.
\end{description}
The values are listed in the last column of Table \ref{comp}. The simple scaling of the mapped area will not be
sufficient if the infrared galaxies have a spatial distribution that is different from the overall cluster
population.  In particular the mapped regions of the three nearer clusters (Abell 1689, Abell 2218, Abell 2390)
are smaller and more confined to the central regions than the two more distant clusters (Abell 370 and Cl
0024+1654).  The infrared sources could preferentially occur in the peripheral regions of clusters where
dynamical interactions between galaxies may be more effective in triggering interactions and bursts of star
formation (Mihos \cite{2004cgpc.symp..278M}).  In this case, the LIRG populations for the complete clusters may
in some cases be less different than the regions currently mapped would make it appear. Future observations with
Spitzer are needed to address this possibility (Werner et al. \cite{spitzer}).

\subsection{LIRGs in Cl 0024+1654 and Abell 370} % 6.1
The clusters Cl 0024+1654 and Abell 370 have comparable redshifts and were observed by ISOCAM with almost
identical observational parameters, such as, on-chip integration time, total area observed in raster mode, and
total observation time.  The total number of cluster sources in Cl 0024+1654 is 13 whereas the corresponding
number is 1 for Abell 370 (Table \ref{comp}).  The observed number of LIRGs in Abell 370 is only 1 whereas 8
were expected from the comparison with Cl 0024+1654.  There is a real difference between the populations of
LIRGs and also in the ratio of the mass to infrared light for the two clusters.  No LIRGs were detected in the
three clusters Abell 1689, Abell 2218 and Abell 2390 at lower redshifts whereas a total number of 3 was expected
from the comparison with Cl 0024+1654.  A population of luminous infrared galaxies is also absent from the
central regions of the Coma and Virgo clusters (Boselli et al. \cite{1997A&A...324L..13B,1998A&A...335...53B};
Quillen et al. \cite{1999ApJ...518..632Q}; Leech et al. \cite{1999MNRAS.310..317L}; Tuffs et al.
\cite{2002ApJS..139...37T,2002ApJS..140..609T}).

%In addition to the comparison between LIRGs in the five clusters, there are two points to be made by
%comparing some of the clusters in a less restrictive manner.
%\begin{enumerate}
%    \item
\subsection{Mid-infrared sources in Abell 1689 and Abell 2218} %6.2
It is, further, interesting to compare the two clusters Abell 1689 and Abell 2218 that are at the same redshift.
The observations of Abell 2218 are more sensitive to 15\,$\mu$m sources than those of Abell 1689 (Table
\ref{comp}). However there is no significant difference between the total number of 15\,$\mu$m sources when
allowance is made for the scanned area and virial mass per unit area. The difference caused by the sensitivity
of the observations is revealed by a comparison of the median fluxes of the 15\,$\mu$m sources which is
$\sim600$ $\mu$Jy for Abell 1689 and $\sim150$ $\mu$Jy for Abell 2218.  All of the 15\,$\mu$m sources in Abell
1689 would have been easily detected in Abell 2218, if present, whereas only one of the sources in Abell 2218
might have been detected in the Abell 1689 measurement. The 15\,$\mu$m sources detected in Abell 2218 have lower
fluxes and luminosities than those in Abell 1689 and this effect is not caused by the sensitivities of the
observations of the two clusters. It is very interesting that Abell 1689 and Abell 2218 follow the same trend
identified in the comparison between Cl 0024+1654 and Abell 370.

Large numbers of cluster sources are detected by ISOCAM at 7\,$\mu$m in both Abell 1689 and Abell 2218. In Abell
2218 the SEDs of most of these 7\,$\mu$m sources are well fit by models of quiescent ellipticals with negligible
SFRs and a median luminosity of $\sim8\times10^{8}$ $\mathrm{L}_\odot$ (Biviano et al. \cite{biviano04}).

\subsection{Dynamical status of clusters and population of mid-infrared sources} %6.3

Cl 0024+1654 is involved in a recent merger, as shown by the two velocity components in Fig. \ref{fig:vel_disp}.
Abell 1689 also has a complex and broad velocity structure, with evidence for three distinct groups that overlap
spatially and are well separated in velocity space (Girardi et al. \cite{1997ApJ...490...56G}). The two clusters
Abell 370 and Abell 2218 show no evidence for major merger activity. The number of luminous 15\,$\mu$m sources
can be, in part, explained by considering the dynamical status of the cluster. Simulations made by Bekki
(\cite{1999ApJ...510L..15B}) show that the time-dependent tidal gravitational field existing in cluster-group
mergers induces secondary starbursts by efficiently transferring large amounts of gas from the disk to the
nucleus. The model establishes a link between the population of starburst or post-starburst galaxies and the
presence of substructures in clusters. The existence of starburst and post-starburst galaxies in a merging
cluster, and spread over a wide area, is an important prediction of this model which seems to be in agreement
with the results from Cl 0024+1654 and Abell 1689.

Furthermore it can be expected that merging clusters at much larger redshifts than Cl 0024+1654 will contain
LIRGs and ULIRGs because the interacting and merging galaxies will be more gas rich (see, e.g. Balogh et al.
\cite{2000ApJ...540..113B}; Kauffmann \& Haehnelt \cite{2000MNRAS.311..576K}; Francis et al.
\cite{2001PASA...18...64F}).

In this context it is interesting to note that Dwarakanath \& Owen (\cite{1999AJ....118..625D}) found different
radio source populations in two very similar clusters. The number density of radio sources in Abell 2125 ($z =
0.246$) exceeded that in \object{Abell 2645} ($z = 0.25$) by almost an order of magnitude. The cluster Abell
2125, with the larger number of radio sources, also shows evidence for a merger.

\subsection{Mid-infrared sources in Cl 0024+1654 and Abell 1689} %6.4

Many infrared sources were found in the cluster Abell 1689 (Fadda et al. \cite{2000A&A...361..827F}) that is at
a much smaller redshift than Cl 0024+1654. No galaxy in Abell 1689 has total infrared luminosity above $1 \times
10^{11}$ L$_{\odot}$ (we use our cosmology and Equation~\ref{eq:elbaz} to convert the 15\,$\mu$m fluxes from
Fadda et al. (\cite{2000A&A...361..827F}) to total infrared luminosities), while half of the 15\,$\mu$m cluster
sources in Cl 0024+1654 are above this luminosity.  All of the 15\,$\mu$m sources detected in Cl 0024+1654 would
have been easily detected in Abell 1689, if present, whereas most of the sources in Abell 1689 would not have
been detected in the observations of Cl 0024+1654 when allowance is made for different sensitivities and
distances. The two clusters have very similar virial radii and masses (Table \ref{comp}). However a much larger
part of the periphery of Cl 0024+1654 was observed because it is at a higher redshift. It is possible that some
of the differences between the 15\,$\mu$m sources arises from a change in the source population between the core
and periphery of the two clusters.

The population of 15\,$\mu$m sources may also be influenced by the history of the cluster. Various processes
such as ram pressure stripping of gas from cluster galaxies, tidal effects and previous starbursts will
inevitably leave less gas in the interacting and merging galaxies to fuel the starburst (e.g. Gunn \& Gott
\cite{1972ApJ...176....1G}; Byrd \& Valtonen \cite{1990ApJ...350...89B}; Fujita \cite{1998ApJ...509..587F}).
These processes may vary from one cluster to another depending on its history and provide dispersion in the
luminosity of the starbursts.  The large number of LIRGs in Cl 0024+1654 are fuelled by the gas rich progenitor
galaxies.  It is conceivable that in Abell 1689 the starbursts are less luminous either because there is a
smaller supply of gas to fuel the outburst, or the luminosity has decreased because a longer time has elapsed
since the last merger or there is a significant difference between the sources in the core and periphery.

The average value of the ratio $\mathrm{SFR}[\mathrm{IR}]/\mathrm{SFR}[\ion{O}{ii}]$ is reasonably comparable
for both clusters (it is $\sim 15$ for Cl 0024+1654 and $\sim11$ for Abell 1689).  The internal properties of
the 15\,$\mu$m sources in Cl 0024+1654 and Abell 1689 are broadly similar even though their distributions of the
luminosities of 15\,$\mu$m sources are different. The colour-magnitude diagrammes of the 15\,$\mu$m sources in
the two clusters are also quite similar.

\section{Conclusions\label{sec:concl}}    % SECTION 7

The cluster Cl 0024+1654 was observed with ISO. A total of 35 sources were detected at 15\,$\mu$m and all have
optical counterparts. Sources with known redshift include four stars, one quasar, three background galaxies, one
foreground galaxy and thirteen cluster galaxies. The remaining 13 sources are likely to be background sources
lensed by the cluster.

The spatial, radial and velocity distributions were obtained for the cluster galaxies. The ISOCAM cluster
galaxies appear to be less centrally grouped (in the cluster) than those not detected at 15\,$\mu$m and the
Kolmogorov-Smirnov test reveals that there is only a 4\% probability that the two distributions are drawn from
the same parent population. No statistically significant differences were found between the velocity
distributions of the 15\,$\mu$m sources and other cluster galaxies in the region mapped by ISOCAM.

Spectral energy distributions were obtained for cluster members and used as indicators of both morphological
type and star forming activity. The ISOCAM sources have as best-fit SEDs predominantly those of spiral or
starburst models observed 1 Gyr after the main starburst.  Star formation rates were computed from the infrared
and the optical data. The SFRs inferred from the infrared are one to two orders of magnitude higher than those
computed from the [\ion{O}{ii}] line emission, suggesting that most of the star forming activity is hidden by
dust.

A colour-magnitude diagramme is given for cluster sources falling within the region mapped by ISOCAM. V and
I-band magnitudes are available for 11 of the cluster sources, and 6 of these have colour properties that are
consistent with Butcher-Oemler galaxies and best-fit SEDs that are typical of spiral models. The remaining
15\,$\mu$m cluster galaxies have colours that are not compatible with Butcher-Oemler galaxies and have best-fit
SEDs that are typical of starburst galaxies 1 Gyr after the main burst. HST images are available for these
latter systems and all have nearby companion galaxies. These results suggest that interactions and mergers are
responsible for some of the luminous infrared sources in the cluster.

The 15\,$\mu$m sources in Cl 0024+1654 were compared with four other clusters observed with ISOCAM.  The results
show that the number of LIRGs in Abell 370 is smaller than expected by about one order of magnitude, if Abell
370 were to be comparable with Cl 0024+1654.  Furthermore no LIRGs were detected in Abell 1689, Abell 2218 and
Abell 2390 when a total of 3 was expected, based on the results from Cl 0024+1654. A comparison of the
mid-infrared sources in Abell 1689 and Abell 2218 shows that the sources in Abell 1689 are more luminous than in
Abell 2218 and follow the same trend identified in the comparison between Cl 0024+1654 and Abell 370. There is
clear evidence for an ongoing merger in Cl 0024+1654 and Abell 1689. The number and luminosities of the
mid-infrared cluster sources seem to be related to the dynamical status and history of the clusters.

\begin{acknowledgements}

{\scriptsize D.C. and B.McB. thank Enterprise Ireland for support. We thank Laura Silva for valuable discussions
regarding the spectral energy distributions. We also thank T. Treu and S. Moran for redshift informations. D.C.
gratefully acknowledges the hospitality of ESA's ISO Data Centre (IDC) at Vilspa, Spain, where part of this work
was completed. JPK acknowledges support from CNRS as well as from Caltech. The ISOCAM data presented in this
paper was analysed using ``CIA", a joint development by the ESA Astrophysics Division and the ISOCAM Consortium.
The ISOCAM Consortium is led by the ISOCAM PI, C. Cesarsky. We thank the anonymous referee for the many comments
that improved the content of the paper.}

\end{acknowledgements}

\begin{appendix}%\appendix

\section{Additional notes on some mid-infrared sources\label{sec:red_des}}

\begin{description}
\item[{\textbf ISO\_Cl0024\_02}] Radio quiet quasar PC 0023+1653 (Schmidt et al. \cite{1986ApJ...306..411S}) at
a redshift of 0.959 with an X-ray luminosity L$_\mathrm{X} = 1.4\times10^{44}$\,erg\,s$^{-1}$ (Soucail et al.
\cite{2000A&A...355..433S}).

\item[{\textbf ISO\_Cl0024\_04}]The 15\,$\mu$m map has two bright regions that may contain contributions from
two optical counterparts. There are no measured redshifts.

\item[{\textbf ISO\_Cl0024\_06}] The 15\,$\mu$m emission appears to have contributions from various galaxies in
the field-of-view. No redshifts are available for these sources.

\item[{\textbf ISO\_Cl0024\_10}] Most of the 15\,$\mu$m emission is centred on a cluster galaxy at redshift $z =
0.400$ (Table~\ref{lw3}), the brightest of the two galaxies in Fig.~\ref{hst2}.  The model fit to the SED is an
Sc (Fig.~\ref{andrea}). The source was also detected in X-rays with ROSAT (B{\" o}hringer et al.
\cite{2000A&A...353..124B}). The optical identification for the ROSAT source is uncertain because there are two
sources in the error box. One of the sources is a typical cluster galaxy while the second is a star-forming
foreground galaxy ($z = 0.2132$). The cluster galaxy is closer to the 15\,$\mu$m coordinates and was adopted as
the mid-infrared counterpart. It is the only cluster AGN detected so far in Cl 0024+1654.

\item[{\textbf ISO\_Cl0024\_16}] The redshift of the optical counterpart on the VLT image is not known. The
15\,$\mu$m emission is centred on one galaxy and partially extends to two very faint optical sources that are
north of the main optical source.

\item[{\textbf ISO\_Cl0024\_17}] The redshift of the optical counterpart on the VLT image has not been measured.
The 15\,$\mu$m emission is centred on the bright optical source and is partially elongated towards two other
sources on the VLT map.

\item[{\textbf ISO\_Cl0024\_18}] The optical counterpart is a large, edge-on, late type, cluster spiral galaxy.
The 15\,$\mu$m emission is centred on the galaxy and has contributions from several other sources. The HST image
reveals a disturbance of the spiral structure.

\item[{\textbf ISO\_Cl0024\_22}] The optical counterpart is a cluster galaxy with $z = 0.3935$. The HST image
(Fig.~\ref{hst2}) reveals a spiral galaxy with an inner ring, a smoother outer arm (Smail et al.
\cite{1997ApJS..110..213S}) and bright knots.

\item[{\textbf ISO\_Cl0024\_23}] The redshift of the optical counterpart is not available. At the 15\,$\mu$m
coordinates, the HST image shows two interacting spiral galaxies and a tidal arm.

\item[{\textbf ISO\_Cl0024\_30}] The source is outside the boundaries of the VLT and HST images
(Fig.~\ref{lw3_fig}) and has a faint optical counterpart (Fig. 4d in Czoske et al. \cite{2001A&A...372..391C}).
\end{description}

\end{appendix}

\end{document}